\documentclass[sigconf]{acmart}

\AtBeginDocument{%
 }

\copyrightyear{2025}
\acmYear{2025}
\setcopyright{cc}
\setcctype{by}
\acmConference[KDD '25]{Proceedings of the 31st ACM SIGKDD Conference on Knowledge Discovery and Data Mining V.2}{August 3--7, 2025}{Toronto, ON, Canada}
\acmBooktitle{Proceedings of the 31st ACM SIGKDD Conference on Knowledge Discovery and Data Mining V.2 (KDD '25), August 3--7, 2025, Toronto, ON, Canada}
\acmDOI{10.1145/3711896.3736967}
\acmISBN{979-8-4007-1454-2/2025/08}

\usepackage{bbm}
\usepackage{balance}
\usepackage{cleveref}
\usepackage{multirow}
\usepackage{graphicx}
\usepackage{enumitem}
\usepackage{xcolor}
\usepackage{algorithm,algpseudocode}%
\usepackage{makecell}
\usepackage{subfigure}

\floatname{algorithm}{Algorithm}

\usepackage{bm}
\usepackage{color}
\usepackage{apxproof}
\usepackage{gensymb}
\usepackage{pifont}%
\usepackage{placeins}  

\usepackage{xspace}
\usepackage{mathtools}

\begin{document}

\newcommand{\ours}{{FlowCF}\xspace}

\title{Flow Matching for Collaborative Filtering}

\author{Chengkai Liu}
\authornotemark[1]
\affiliation{
  \institution{Texas A\&M University}
  \city{College Station, TX}
  \country{USA}}
\email{liuchengkai@tamu.edu}

\author{Yangtian Zhang}
\authornotemark[1]
\affiliation{
  \institution{Yale University}
  \city{New Haven, CT}
  \country{USA}}
\email{yangtian.zhang@yale.edu}

\author{Jianling Wang}
\affiliation{
  \institution{Google Deepmind}
  \city{Mountain View, CA}
  \country{USA}}
\email{jianlingw@google.com}

\author{Rex Ying}
\affiliation{
  \institution{Yale University}
  \city{New Haven, CT}
  \country{USA}}
\email{rex.ying@yale.edu}

\author{James Caverlee}
\affiliation{
  \institution{Texas A\&M University}
  \city{College Station, TX}
  \country{USA}}
\email{caverlee@cse.tamu.edu}

\begin{abstract}
Generative models have shown great promise in collaborative filtering by capturing the underlying distribution of user interests and preferences. However, existing approaches struggle with inaccurate posterior approximations and misalignment with the discrete nature of recommendation data, limiting their expressiveness and real-world performance. 
To address these limitations, we propose \textbf{\ours}, a novel flow-based recommendation system leveraging flow matching for collaborative filtering. 
We tailor flow matching to the unique challenges in recommendation through two key innovations: (1) a \textit{behavior-guided prior} that aligns with user behavior patterns to handle the sparse and heterogeneous user-item interactions, and (2) a \textit{discrete flow framework} to preserve the binary nature of implicit feedback while maintaining the benefits of flow matching, such as stable training and efficient inference.
Extensive experiments demonstrate that \ours achieves state-of-the-art recommendation accuracy across various datasets with the fastest inference speed, making it a compelling approach for real-world recommender systems. The code is available at \url{https://github.com/chengkai-liu/FlowCF}.

\end{abstract}



\maketitle

\section{Introduction}

\begin{figure}[t]
    \includegraphics[width=0.48\textwidth]{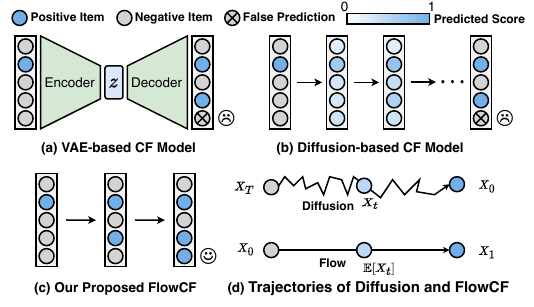}
    \caption{Illustration of VAE and diffusion-based CF models, the proposed \ours, along with a trajectory comparison between diffusion process and flow in \ours.}
    \label{fig:intro_example}
\end{figure}

Collaborative filtering (CF)~\cite{schafer2007collaborative, koren2021advances} -- as the foundational component of model-based recommender systems~\cite{liu2024mamba4rec, liu2024behavior} -- aims to recommend items to users by leveraging the preferences of similar users based on their past interactions with items. One exciting thread is the rise of \textit{generative collaborative filtering}, ranging from variational autoencoders (VAEs) \cite{kingma2013auto} to more advanced diffusion models~\cite{ho2020denoising, song2020score}. These generative CF approaches model the distribution of user preferences to generate diverse and realistic interaction patterns, thus better handling challenges like noisy data~\cite{wang2021denoising} and cold-start scenarios~\cite{chen2022generative}, in comparison with non-generative methods that focus on learning user and item representations for prediction~\cite{he2017neural, he2020lightgcn}.

Early generative CF methods like VAE-based methods~\cite{liang2018variational, ma2019learning, shenbin2020recvae} pioneered the approximation of user preference distributions, demonstrating the potential of generative modeling for recommendation (Figure~\ref{fig:intro_example}(a)). However, these approaches often struggle with inaccurate posterior approximations. This inaccuracy can blur the reconstructed user preferences, ultimately hindering recommendation accuracy. Furthermore, their reliance on variational approximation and Kullback-Leibler (KL) divergence often leads to unstable training and slow convergence. 
DiffRec~\cite{wang2023diffusion} is the leading example of using diffusion models in CF. Although it solves the aforementioned issues of VAE-based methods during training, the stochastic nature~\cite{song2020score} of its underlying Gaussian diffusion processes~\cite{sohl2015deep, ho2020denoising} results in computational inefficiency with multiple sampling steps during inference, as illustrated in Figure~\ref{fig:intro_example}(b). 
More fundamentally, the discrete nature of user-item interactions and binary implicit feedback in recommendations poses a challenge. The reliance on a continuous diffusion process applying Gaussian noise to these discrete interactions within a continuous state space risks distorting personalized preferences. This mismatch can compromise diffusion models' ability to accurately capture sparse and heterogeneous patterns inherent in user-item interactions, thereby limiting their effectiveness in learning user preference distributions.

To address these limitations, we propose leveraging the \textit{flow matching} framework~\cite{lipman2022flow} for generative CF. Flow matching learns continuous normalizing flows~\cite{chen2018neural} by modeling the vector field that transforms a simple prior distribution into a complex target distribution, such as user preferences. Unlike stochastic diffusion processes with curved probability paths, flow matching enables the learning of deterministic, straight flow trajectories. These trajectories offer both stable training and efficient sampling, as fewer steps are needed for straight paths compared to curved paths. Furthermore, the framework's flexibility in designing makes flow matching particularly well-suited for designing advanced generative recommendation models. These models not only achieve superior recommendation performance but also meet the requirement of fast inference speed, which is essential for real-world recommender systems.

Despite the advantages of flow matching, its application to collaborative filtering presents unique challenges. A key component of flow matching is the prior distribution, which defines the initial state of the generative process. While Gaussian priors are widely adopted in many generative tasks~\cite{liu2022flow, albergo2022building}, they may not be suitable for real-world recommendation data, which often exhibits sparsity and popularity bias. 
Additionally, the continuous normalizing flows used in flow matching struggle to naturally simulate the discrete transitions inherent in recommendation scenarios. 

To this end, in this work, we introduce \textbf{\ours}, a novel flow-based recommendation model that adapts flow matching to the unique characteristics of collaborative filtering through two key innovations: a \textit{behavior-guided prior} and a \textit{discrete flow framework}. 
Firstly, to better capture user behavior patterns, we introduce a behavior-guided prior, which is designed to align closely with real-world interactions. 
This prior is a probability distribution derived from the global frequencies of interacted items, which inherently reflects underlying user preferences. 
By incorporating domain-specific knowledge into the learning process, the behavior-guided prior ensures that the flow training process is guided by realistic behavioral data, leading to more effective and meaningful model training.
Secondly, we propose a discrete flow framework that explicitly models the binary nature of implicit feedback.
While this framework respects the inherent discreteness of the data, it maintains a linear evolution of mathematical expectations between the source and target distributions, as illustrated in Figure~\ref{fig:intro_example}(d). 
This design bridges the gap between discrete interactions and continuous state spaces, providing a robust theoretical foundation for our approach. Importantly, \ours retains the benefits of flow matching, such as training stability and efficient sampling. The straight flow trajectories of \ours (Figure~\ref{fig:intro_example}(d)) enable faster inference speed with fewer sampling steps compared to the curved trajectories of diffusion-based CF models, ensuring that our discrete flow framework delivers superior recommendation performance with rapid inference capabilities.

In summary, our key contributions are as follows:
\begin{itemize}
[leftmargin=*,noitemsep,topsep=1.5pt]
    \item We propose \ours, a novel flow-based recommendation model using flow matching that advances the generative paradigm for collaborative filtering.
    \item We adapt flow matching to recommendation scenarios via a behavior-guided prior capturing user behavior patterns and a discrete flow framework modeling binary implicit feedback.
    \item We conduct comprehensive experiments on multiple datasets, demonstrating that \ours achieves state-of-the-art performance, with both stable training and fast inference speed, addressing the practical needs of real-world recommender systems.
\end{itemize}

\section{Preliminaries}

\subsection{Generative Collaborative Filtering}
We denote the set of users as $\mathcal{U}=\{u_i\}_{i=1}^{|\mathcal{U}|}$, the set of items as $\mathcal{I}=\{i_i\}_{i=1}^{|\mathcal{I}|}$, and the user-item interaction matrix as $\bar X \in \{0, 1\}^{|\mathcal{U}| \times |\mathcal{I}|}$. Recommendation systems aim to predict the missing entries in $\bar X$ effectively, uncovering unobserved user preferences and potential interactions. CF approaches this problem by recommending items to users based on the interaction patterns of other users with similar interests and preferences.

Generative CF offers a different paradigm compared to traditional methods that rely on latent representations for users and items~\cite{rendle2012bpr, he2017neural, he2020lightgcn}. Its core principle is to model probability distributions,
specifically learning a transformation from the source distribution of historical interactions $p_{\text{data}}$ to a target distribution $q_{\text{pref}}$ that reflects user preferences via probabilistic generative models. By interpreting $\bar X$ as a sample from $p_{\text{data}}$ and the predicted interaction matrix $\hat X$ as a sample from $q_{\text{pref}}$, the process can be formalized as follows:
\begin{equation}
    \begin{aligned}
        &\bar X \sim p_{\text{data}}, \\
        &q_{\text{pref}} = f_\theta(p_{\text{data}}),  \\
        &\hat X \sim q_{\text{pref}},
    \end{aligned}
\end{equation}
where $f_\theta$ is a learnable non-linear function with parameters $\theta$ that transforms the distribution. The transformation function aims to effectively reconstruct and predict user-item interactions. From a probabilistic perspective, generative CF learns a mapping from the source sample $\bar X$ to the target $\hat X$ through a learned transformation.

\begin{figure*}[ht]
    \centering
    \includegraphics[width=0.86\textwidth]{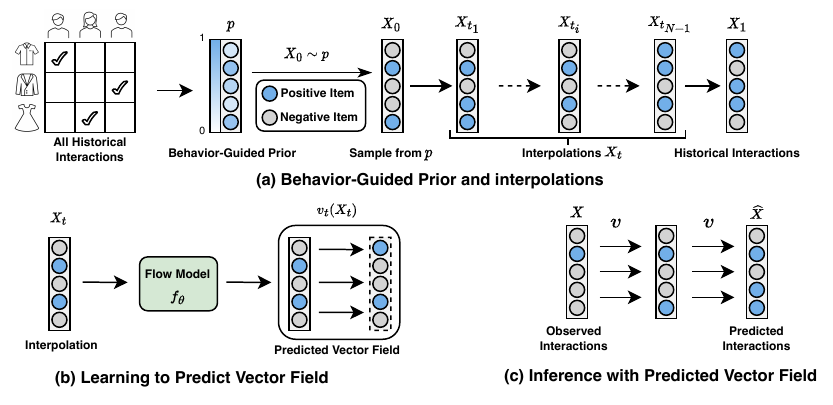}
    \caption{Overview of \ours. (a) A set of interpolations $X_t$ is created between the sample $X_0$ from the behavior-guided prior and the historical interactions $X_1$. (b) During training, \ours learns to predict the vector field using a flow model approximating the true flow. (c) \ours leverages the predicted vector field from the flow model for inference.}
    \label{fig:framework}
\end{figure*}

\subsection{Flow Matching}

Flow matching~\cite{lipman2022flow, liu2022flow} is a technique to train deep flow-based generative models, which model probability distributions by leveraging normalizing flow~\cite{chen2018neural, papamakarios2021normalizing}. Flow matching can be formulated as learning a time-dependent vector field $v_t(x)$ that transforms a source distribution $p$ to a target distribution $q$, and solving the \textbf{flow matching problem}:
\begin{equation}
    \text{Find } v_t(x) \text{ generating the probability path } p_t,
\end{equation}
where $p_0 = p$, $p_1 = q$, and $t \in [0, 1]$.
The flow matching method defines a time-dependent flow $\psi_t(x)$ as an ordinary differential equation (ODE):
\begin{equation}
    \frac{\mathrm d \psi_t(x)}{\mathrm d t} = v_t(\psi_t(x)),
\end{equation}
where $\psi_0(x) = x$ is an initial condition, and $v_t(x)$ is the target vector field known to generate the desired probability path $p_t$. The vector field $v$ generates the probability path $p_t$ if its flow $\psi$ satisfies:
\begin{equation}
    \label{eq:flow}
    X_t = \psi_t(X_0) \sim p_t,
\end{equation}
where $X_0 = \psi(X_0, 0) \sim p$. The goal of flow matching is to learn a vector field $v$ such that its flow $\psi$ such that the target samples $X_1$ can be obtained from the source samples $X_0$:
\begin{equation}
    X_1 = \psi_1(X_0) \sim q.
\end{equation}
The objective of flow matching is to learn the vector field $v$ such that it closely approximates the true flow and the ground truth vector field $u$. Therefore, the flow matching loss is:
\begin{equation}
    \label{eq:fm_loss}
    \mathcal{L}_{\mathrm{FM}}(\theta) = \mathbb{E}_{t, X_t} \|v_t(X_t) - u_t(X_t)\|^2.
\end{equation}
An objective of conditional flow matching (CFM) proposed in~\cite{lipman2022flow} has been proved to have identical gradients and result in the same optima as the original objective in Equation~\ref{eq:fm_loss}:
\begin{equation}
    \label{eq:cfm_loss}
    \mathcal{L}_{\mathrm{CFM}}(\theta) = \mathbb{E}_{t, X_1, X_t} \|v_t(X_t) - u_t(X_t \mid X_1)\|^2.
\end{equation}
Flow matching combines ideas from generative modeling and optimal transport, making it conceptually related to diffusion models but distinct in its use of explicit interpolation paths. One of the key advantages of flow matching over traditional diffusion models is that it can potentially learn straighter, more efficient probability paths between the source and target distributions. This can lead to faster sampling with fewer sampling steps and potentially samples with better quality. Moreover, flow matching provides a unified framework that encompasses both flows and stochastic diffusion processes, allowing for a deeper understanding of these generative modeling approaches.

\section{Flow-Based Recommender -- \ours}
Flow matching has shown promise in various domains due to its ability to learn complex data distributions with efficient sampling. However, applying flow matching to collaborative filtering presents several distinct challenges. Unlike conventional generative tasks, which assume a simple Gaussian prior, recommendation data is inherently sparse and heterogeneous, requiring a more expressive prior distribution for training. 
Furthermore, user-item interactions are inherently discrete, making it insufficient to directly apply continuous flow models without adaptation. 
Finally, practical recommender systems require both efficient training and fast inference.  Therefore, we must ensure that our design leverages the benefits of flow matching without sacrificing training and inference efficiency. 
These challenges motivate our investigation into the following research questions:
\begin{itemize}
 [leftmargin=*,noitemsep,topsep=1.5pt]
     \item \textbf{RQ1}: How can we design an appropriate prior distribution for flow matching in recommendations?
     \item \textbf{RQ2}: How can we effectively adapt flow models to handle discrete and binary implicit feedback data?
     \item \textbf{RQ3}: How can we train and optimize \ours and learn to predict the vector field effectively?
     \item \textbf{RQ4}: What factors contribute to the fast inference speed and swift recommendation response of \ours?
\end{itemize}
 
\subsection{Overall Framework}
To harness the generative capabilities and flexibility of flow matching, we propose \ours, a flow-based recommender system that models the evolution of user-item interactions from a noisy source distribution to a target preference distribution.
\ours employs a flow model $f_\theta$, implemented as a multi-layer perceptron (MLP), which learns the vector field governing the flow dynamics. Flow matching is utilized to facilitate the accurate approximation of the flow by aligning the learned transformations with the underlying user interaction patterns.

As illustrated in Figure~\ref{fig:framework}(a), the overall process begins with sampling $X_0$ from a selected prior distribution $p$.
A set of interpolations $X_t$ is then defined along the flow trajectory in Equation~\ref{eq:flow}, connecting $X_0$ to the historical user-item interactions $\bar X$, which serve as the sample $X_1$ from the target distribution. As shown in Figure~\ref{fig:framework}(b), the flow model $f_\theta$ learns the vector field to transform $X_0$ into $X_1$, producing an approximation of the true interaction distribution. The objective is to minimize the discrepancy between the predicted vector field $v$ and the ground truth vector field $u$ in Equation~\ref{eq:cfm_loss}. Once trained, \ours can use the flow model to predict vector fields for inference, as depicted in Figure~\ref{fig:framework}(c).

Specifically, to construct these interpolations $X_t$ at different time steps for training, discretize the time interval into $N$ steps:
\begin{equation}
\label{eq:steps}
T=\left\{t_i \left\lvert\, t_i=\frac{i}{N}\right., \quad i=0,1, \ldots, N\right\},
\end{equation}
where $t_i \in [0, 1]$, starting from $t_0 = 0$ to $t_N = 1$.
This discretization allows for the numerical simulation of the flow using the Euler method with a constant step size of $1 / N$. 
By encoding the step $t$ and integrating step embeddings via concatenation into the flow model $f_\theta$, the flow model learns step-dependent transformations and predicts the vector field $v_t$ at each step $t$.

A critical aspect of the \ours framework is the selection of an appropriate prior distribution $p$ and the definition of flow dynamics tailored for recommendation tasks. These components are addressed in detail in the subsequent sections.

\subsection{Behavior-Guided Prior (RQ1)}
Flow matching typically relies on a prior distribution for training. In many generative tasks, such as image generation~\cite{lipman2022flow, liu2022flow}, the Gaussian distribution $p = \mathcal{N} (0, I )$ is commonly employed due to its mathematical tractability and universality. However, in collaborative filtering (CF), the distribution of interactions exhibits distinct characteristics. Specifically, the user-item interactions are highly sparse and often follow a long-tailed popularity bias. Furthermore, the role of our \ours is to transform a noisy user behavior distribution -- rather than a pure noise like a Gaussian distribution -- into the target distribution. Consequently, a Gaussian prior is ill-suited for flow matching in CF, as it fails to capture the inherent sparsity of interactions and the dynamic evolution of user behaviors. This misalignment between the prior and the true requirements of CF can lead to insufficient learning and suboptimal performance.

To address this challenge, we hypothesize that \textit{using a prior distribution which is close to the true user behavior distribution is beneficial to effective training of flow matching models in recommendation tasks}. Therefore, we propose that a prior distribution derived from real-world user behaviors can better guide the flow matching process. Such a behavior-guided prior enables the flow model to learn more meaningful transitions from the prior to the distribution of historical interactions. By incorporating the inherent sparsity of user-item interactions and global item popularity -- a critical factor in recommendation -- this approach ensures a more accurate representation of the underlying user behavior distribution.

To operationalize this idea, we design a \textit{Behavior-Guided Prior Distribution} that leverages the global item frequency vector $\mathbf f \in \mathbb R^{|\mathcal I|}$. Let $\mathrm f_i$ denote the frequency of item $i$ across all historical user-item interactions $\bar X$:
\begin{equation}
\mathrm f_i=\frac{\text{number of interactions for item } i}{\text{total number of users } |\mathcal U|}.
\end{equation}
This frequency vector $\mathbf f$ captures the global popularity of items across all historical user-item interactions, providing a meaningful foundation for constructing the behavior-guided prior distribution.

To model the prior, we employ the Bernoulli distribution, which is well-suited for binary outcomes. Specifically, the Bernoulli distribution models whether an event occurs (1) or does not occur (0) based on a given probability. This aligns with our goal of representing user-item interactions as binary implicit feedback, where 1 indicates positive feedback and 0 indicates the absence of such feedback.
The behavior-guided prior distribution is then constructed by applying the Bernoulli distribution with expanded global item frequency $\mathbf f$:
\begin{equation}
    \label{behavior_guided_prior}
    p = \mathrm{Bernoulli}(\mathbf 1_{|\mathcal U|} \otimes \mathbf f) .
\end{equation}
This ensures that the behavior-guided prior distribution aligns with the historical item frequencies across all users, thereby providing a more suitable starting point for the flow matching training process.
The starting point $X_0$ for the flow matching training process is sampled from the behavior-guided prior distribution:
\begin{equation}
    \label{eq:sample_x0}
X_0 \sim  \mathrm{Bernoulli}(\mathbf 1_{|\mathcal U|} \otimes \mathbf f) \in \{0, 1\}^{|\mathcal U| \times |\mathcal I|},
\end{equation}
where $X_0$ comprises only 0 and 1 values, representing negative and positive implicit feedback, respectively. The sample $X_0$ remains behavior-guided, reflecting the global item frequencies, as well as the sparsity and popularity bias inherent in user-item interactions.

\subsection{Discrete Flow Framework (RQ2)}
Traditional flow matching and diffusion-based DiffRec operate in a continuous state space. In continuous flow matching, the flow trajectory between the source sample $X_0$ and target sample $X_1$ is often defined using linear interpolation~\cite{liu2022flow}. This interpolation can be defined as:
\begin{equation}
    \label{eq:interpolation_continuous}
    X_t = t X_1 + (1 - t) X_0 \sim p_t,
\end{equation}
where $X_t$ represents a linear interpolation between $X_0$ and $X_1$. This continuous linear interpolation ensures a smooth transition between the source and target distributions. While there are alternative interpolation methods, such as the sinusoidal interpolation~\cite{albergo2022building}, the straight flow trajectory achieved via linear interpolation aligns well with recommendation tasks and is more computationally efficient compared to nonlinear or stochastic interpolations.

However, in collaborative filtering, we primarily model rich implicit feedback (e.g., clicks or purchases) from users, which is inherently binary and discrete, as opposed to the rarer explicit feedback. Given that both the historical interactions $X_1$ and the sample $X_0$ (Equation~\ref{eq:sample_x0}) from the behavior-guided prior are binary, preserving this binary structure poses a significant challenge.
Continuous flow matching fails to preserve the discrete structure of interaction data because continuous interpolations, such as Equation~\ref{eq:interpolation_continuous}, produce fractional values that do not align with valid binary interactions. Such approximations to discrete interactions can lead to inaccuracies in recommendations. 
To address this limitation, we propose a \textit{discrete flow framework} that operates in a discrete state space~\cite{campbell2024generative, gat2024discrete}, explicitly modeling binary implicit feedback, along with a series of discretization strategies, to ensure the integrity of the interactions.

\vspace{3pt}
\noindent \textbf{Discretized Linear Interpolation.}
To preserve the discrete nature of the data, we propose a discretized linear interpolation:
\begin{equation}
\label{eq:interpolation_discrete}
X_t=M_t \odot X_1 + (\mathbf 1_{|\mathcal U| \times |\mathcal I|} - M_t) \odot X_0.
\end{equation}
where $M_t \in \{0, 1\}^{|\mathcal U| \times |\mathcal I|}$ is a binary mask with each element $M_t^i \sim \operatorname{Bernoulli}(t)$ and $\odot$ denotes element-wise multiplication. This formulation ensures that each entry $X_t^i$ is a mixture of the historical interaction $X_1^i$ and the source sample $X_0^i$, with a probability $t$:
\begin{equation}
X_t^i= 
\begin{cases}
X_1^i & \text{with probability } t, \\
X_0^i & \text{with probability } 1-t.
\end{cases}
\end{equation}
This discretized linear interpolation maintains the binary structure of the interaction data, ensuring that $X_t$ remains step-dependent for all steps $t \in T$. Despite the discretization, which restricts transitions to binary values, the expectation of $X_t$ still behaves like a smooth linear function:
\begin{equation}
    \label{eq:expectation}
    \mathbb E[X_t] = t X_1 + (1 - t) X_0.
\end{equation}
Therefore, despite the use of discrete variables, the discrete flow framework maintains a theoretical alignment with continuous interpolation in expectation.

\vspace{3pt}
\noindent \textbf{Vector Field of Expectation.}
The flow dynamics under the framework are governed by the vector field of expectation  $u_t(\mathbb E[X_t] \mid X_1)$, which describes the rate of change of the expectation $\mathbb E[X_t]$. This vector field is derived from Equation~\ref{eq:expectation} as an ODE:
\begin{equation}
    \label{eq:vector_field1}
    u_t(\mathbb E[X_t] \mid X_1) = \frac{\mathrm d \mathbb E[X_t]}{\mathrm d t} = X_1 - X_0.
\end{equation}
Here, the difference vector $X_1 - X_0$ represents the direction of the flow, guiding the transition from $X_0$ to $X_1$.
By substituting $X_0 = \frac{\mathbb E[X_t] - t X_1}{1 - t}$, the vector field can be reformulated as:
\begin{equation}
    \label{eq:vector_field2}
    u_t(\mathbb E[X_t] \mid X_1) = \frac{X_1 - \mathbb E[X_t]}{1 - t}.
\end{equation}
This formulation mirrors the continuous flow vector field but is adapted for discrete data. While the vector field is defined in terms of the expectation $\mathbb E[X_t]$, the actual updates are applied to the discrete $X_t$, ensuring that the binary structure of interactions is preserved.
Similarly, the predicted vector field $v_t(\mathbb E[X_t])$ is computed as:
\begin{equation}
    v_t(\mathbb E[X_t]) = \frac{\hat X_1 - \mathbb E[X_t]}{1 - t},
\end{equation}
where $\hat X_1$ is the predicted user interactions.

In summary, we propose a discrete flow framework designed for recommendations, incorporating the discrete sample from behavior-guided prior and discretized linear interpolation to preserve the binary nature of implicit feedback. By leveraging a vector field of expectation, we bridge the gap between discrete and continuous state spaces, ensuring theoretical consistency while maintaining the discrete structure of interactions. This framework enables smooth and interpretable transitions from source to target distributions, addressing the unique challenges of collaborative filtering.

\subsection{Training \ours (RQ3)}
The training process for \ours focuses on optimizing the flow model to accurately approximate the vector field to transform user-item interactions. This optimization can be interpreted as aligning a predicted vector field of expectation with a ground truth vector field of expectation. By substituting the ground truth vector field $u$ and the predicted vector field $v$ into Equation~\ref{eq:cfm_loss}, we derive the following loss function:
\begin{equation}
    \mathcal L_t = \mathbb{E}_{t, X_1, X_t} \left[\left\|\frac{\hat X_1 - \mathbb E[X_t]}{1 - t} - \frac{X_1 - \mathbb E[X_t]}{1 - t}\right\|^2 \right],
\end{equation}
which encourages $\hat X_1$ to accurately predict the target interaction matrix $X_1$.
Consequently, the flow model $f_\theta$ can be optimized to minimize the mean squared error between its predictions and the true interactions, leading to the following simplified loss:
\begin{equation}
\label{eq:our_loss}
	\mathcal{L}_t = \mathbb{E}_{t, X_1, X_t} \left[\|\hat X_1 - X_1\|^2 \right],
\end{equation}
where $X_t$ can be viewed as the noisy interaction matrix at step $t$, and $X_1$ is the target interaction matrix. Through this formulation, the flow model $f_\theta$ can be used to directly predict the target interactions $\hat X_1$ and subsequently used to compute the predicted vector field $v$ with $\hat X_1$. This alternative approach, given by
\begin{equation}
    \hat X_1 = f_\theta(X_t, t),
\end{equation}
eliminates the need for explicit vector field estimation, which is often more challenging. Instead, it focuses on capturing the underlying structure of user-item interactions, thereby improving the ability to generate accurate recommendations.

Unlike KL divergence-based approaches~\cite{liang2018variational, ma2019learning,shenbin2020recvae}, which often suffer from issues such as mode collapse or over-penalization of low-density regions in the learned distribution, \ours directly optimizes the interaction predictions. This approach avoids reliance on distributional matching and better captures the structure of user-item relationships. The training process is detailed in Algorithm~\ref{algo:training}.

\begin{algorithm}[H]
	\caption{\textbf{Training \ours}}  
	\label{algo:training}
	\begin{algorithmic}[1]
		\Require All user interactions $\Bar X$ and the randomly initialized parameters $\theta$ of the flow model $f_\theta$
            \Repeat 
            \State Sample a batch of users $\mathcal U_b$ and their interactions as $X_1 \subset \bar X$
            \State Create discrete steps $T$ by Equation~\ref{eq:steps}
            \State Sample a batch of steps $t$ from $T^{|\mathcal U_b|}$
            \State Sample $X_0$ from the behavior-guided prior 
            \State Compute the interpolation $X_t$ by Equation~\ref{eq:interpolation_discrete}
            \State Predict $\hat X_1 = f_\theta(X_t, t)$
            \State Compute loss $\mathcal L_t$ by Equation~\ref{eq:our_loss}
            \State Update $\theta$ via gradient descent on $\nabla_\theta\mathcal L_t$
            \Until{Converged}
            \Ensure Optimized parameters $\theta$ of the flow model $f_\theta$.
	\end{algorithmic}
\end{algorithm}
\setlength{\textfloatsep}{0.28cm}

\begin{algorithm}[H]
	\caption{\textbf{Inference with \ours}}  
	\label{algo:inference}
	\begin{algorithmic}[1]
		\Require Trained flow model $f_\theta$ with parameters $\theta$, observed user interactions $X$, number of discretization steps $N$
        \State Set the starting step $s$ 
        \State Initialize $X_t \gets X$
        \For{$i = s \text{ to } N-2$}
        	\State Set the current step $t \gets t_i$
            \State Predict $\hat{X} = f_\theta(X_t, t)$ 
            \State Compute the predicted $v_t \gets (\hat{X} - X_t) / (1 - t)$
            \State Update $X_t$ by Equation~\ref{eq:update_xt}
            \State Preserve observed interactions $X_t \gets X_t \vee X$
        \EndFor
        \State Predict $\hat X = f_\theta(X_t, t)$ in the last step $t = t_{N-1}$
        \Ensure Predicted interactions $\hat X$
	\end{algorithmic}
\end{algorithm}

\subsection{Inference with \ours (RQ4)}
During the inference process, the observed user interactions $X$ can be treated as a noisy interaction matrix, analogous to an interpolation $X_t$ between prior sample $X_0$ from the behavior-guided prior and the ground truth interactions $X_1$. To initiate the inference process, a starting sampling step needs to be selected. Since the starting point $X$ is not purely noise but contains meaningful information about user historical behavior, it is unnecessary to begin at $t = 0$. Instead, the inference process can start from a later step, significantly reducing the number of sampling steps required.\
To preserve the binary nature of the interactions $X_t$ throughout the inference process, \ours employs the following update rule, where each update corresponds to an element-wise argmax over the predicted field, to compute $X_{t+\frac{1}{N}}$ at the next step of $t$:
\begin{equation}
\label{eq:update_xt}
X_{t+\frac{1}{N}}^i= 
\begin{cases}
1 & \text{if } X_t^i + \frac{v_t^i}{N} \geq 0.5, \\
0 & \text{otherwise}.
\end{cases}
\end{equation}
where $v_t$ is the predicted vector field, and $1/N$ is the step size that controls the granularity of the discrete steps. 
The complete inference process is described in Algorithm~\ref{algo:inference}.

\begin{table*}[t]
\caption{Overall performance comparison. The best results of \ours are highlighted in bold, and the best baseline results are underlined. The relative improvements of \ours over the best baselines are indicated as Improve. The symbol $*$ indicates the improvement over the best baseline is statistically significant with $p$-value < 0.05.}
\label{tab:overall}
\begin{tabular}{l|cccc|cccc|cccc}
\hline
& \multicolumn{4}{c|}{\textbf{MovieLens-1M}} & \multicolumn{4}{c|}{\textbf{MovieLens-20M}} & \multicolumn{4}{c}{\textbf{Amazon-Beauty}} \\ \hline
\textbf{Methods} & \textbf{R@10} & \textbf{R@20}  & \textbf{N@10}  & \textbf{N@20}  & \textbf{R@10}  & \textbf{R@20}  & \textbf{N@10}  & \textbf{N@20}  & \textbf{R@10}  & \textbf{R@20}  & \textbf{N@10}  & \textbf{N@20}  \\ \hline
\textbf{MF-BPR} & 0.1764 & 0.2672 & 0.1924 & 0.2124 & 0.1980 & 0.2918 & 0.1779 & 0.2023 & 0.0871 & 0.1259 & 0.0487 & 0.0588 \\
\textbf{LightGCN} & 0.1859 & 0.2753 & 0.1849 & 0.2158 & 0.2055 & 0.3020 & 0.1855 & 0.2103 & 0.1032 & 0.1482 & 0.0579 & 0.0696 \\
\textbf{SGL} & 0.1936 & 0.2900 & 0.2086 & 0.2301 & 0.2219 & 0.3180 & 0.1947 & 0.2247 & 0.1093 & 0.1566 & 0.0624 & 0.0746 \\
\hline
\textbf{CDAE} & 0.1703 & 0.2517 & 0.1870 & 0.2045 & 0.2047 & 0.2998 & 0.2180 & 0.1936 & 0.0844 & 0.1285 & 0.0511 & 0.0614 \\
\textbf{Mult-DAE} & 0.1797 & 0.2667 & 0.1899 & 0.2106 & 0.2431 & 0.3488 & 0.2114 & 0.2399 & 0.0949 & 0.1388 & 0.0551 & 0.0666 \\
\textbf{Mult-VAE} & 0.1855 & 0.2751 & 0.1952 & 0.2160 & 0.2472 & 0.3531 & 0.2134 & 0.2424 & 0.1009 & 0.1417 & 0.0583 & 0.0690 \\
\textbf{MacridVAE} & 0.1775 & 0.2593 & 0.1903 & 0.2082 & 0.2417 & 0.3399 & 0.2153 & 0.2408 & 0.1147 & 0.1524 & 0.0695 & 0.0794 \\ 
\textbf{RecVAE} & 0.2049 & \underline{0.3047} & 0.2146 & 0.2368 & 0.2487 & \underline{0.3538} & 0.2217 & 0.2491 & 0.1153 & \underline{0.1584} & 0.0677 & 0.0790 \\
\textbf{DiffRec} & \underline{0.2055} & 0.3030 & \underline{0.2208} & \underline{0.2423} & \underline{0.2530} & 0.3516 & \underline{0.2362} & \underline{0.2594} & \underline{0.1190} & 0.1570 & \underline{0.0731} & \underline{0.0832} \\
\hline
\textbf{\ours} & \textbf{0.2157}* & \textbf{0.3145}* & \textbf{0.2320}* & \textbf{0.2530}* & \textbf{0.2580}* & \textbf{0.3585}* & \textbf{0.2394}* & \textbf{0.2634}* & \textbf{0.1245}* & \textbf{0.1633}* & \textbf{0.0755}* & \textbf{0.0856}* \\ 
\textbf{\% Improve.} & 4.96\% & 3.35\% & 5.07\% & 4.42\% & 1.98\% & 1.33\% & 1.02\% & 1.20\% & 4.62\% & 3.10\% & 3.28\% & 2.88\% \\ \hline

\end{tabular}
\end{table*}

In practice, \ours achieves superior performance within just two sampling steps, starting from the second-to-last step during inference. Compared to diffusion models, \ours leverages flow matching, which benefits from straight flow trajectories and eliminates the need to start from $t=0$ with unnecessary sampling steps. This property enables efficient sampling and allows \ours to require much fewer sampling steps, resulting in faster inference speeds and facilitating quicker recommendations.

\section{Experiments}

\subsection{Experimental Setup}
\noindent \textbf{Datasets.} To evaluate our proposed model, we conduct our experiments on three public datasets in real-world scenarios: \textbf{MovieLens-1M}~\cite{harper2015movielens}, \textbf{MovieLens-20M}, and \textbf{Amazon-Beauty}~\cite{mcauley2015image}. These datasets vary in scale and density, which allows us to evaluate the performance of the models across different levels of sparsity. Although they consist of explicit ratings, we have intentionally chosen them to investigate the performance of learning from implicit feedback. To convert explicit ratings into implicit feedback, we binarize the ratings by treating ratings of four and above as positive interactions and discarding ratings < 4. We split the dataset into training, validation, and test sets at a ratio of 8:1:1. To be consistent with the preprocessing in previous studies~\cite{liang2018variational}, we exclude users and items with fewer than five interactions from the datasets. Table~\ref{tab:dataset} provides a summary of the statistics of the datasets after preprocessing.

\begin{table}[ht]
\normalsize
  \centering
  \caption{Statistics of the experimented datasets.}
  \label{tab:dataset}
  \resizebox{0.47\textwidth}{!}{%
  \begin{tabular}{lcccc}
    \toprule
    \multicolumn{1}{l}{\textbf{Dataset}} & \multicolumn{1}{c}{\textbf{\# Users}} & \multicolumn{1}{c}{\textbf{\# Items}} & \multicolumn{1}{c}{\textbf{\# Interactions}} & \multicolumn{1}{c}{\textbf{Density}} \\ 
    \midrule
    MovieLens-1M & 6,034 & 3,126 & 574,376 & 3.04\%\\
    MovieLens-20M & 136,674 & 13,681 & 9,977,451 & 0.53\%\\
    Amazon-Beauty & 10,553 & 6,087 & 94,148 & 0.15\%\\
  \bottomrule
\end{tabular}}
\end{table}

\noindent \textbf{Baselines.} We compare \ours with several baselines, including non-generative methods like \textbf{MF-BPR}~\cite{rendle2012bpr}, \textbf{LightGCN}~\cite{he2020lightgcn}, and \textbf{SGL}~\cite{wujc2021self}; DAE-based methods like \textbf{CDAE}~\cite{wu2016collaborative} and \textbf{Mult-DAE}~\cite{liang2018variational}; VAE-based methods like \textbf{Mult-VAE}~\cite{liang2018variational}, \textbf{MacridVAE}~\cite{ma2019learning}, and \textbf{RecVAE}~\cite{shenbin2020recvae}; and the diffusion-based method \textbf{DiffRec}~\cite{wang2023diffusion}. The detailed descriptions of baselines are in Appendix~\ref{sec:baseline}.

\vspace{1pt}
\noindent \textbf{Evaluation Metrics.} We use two standard metrics, Recall@$K$ (R@$K$) and NDCG@$K$ (N@$K$), following the all-ranking protocol~\cite{he2020lightgcn}, which ranks all non-interacted items for each user. For each metric, we evaluate the top-$K$ items, with $K \in [10, 20]$.

\vspace{1pt}
\noindent \textbf{Implementation Details.} We use 9 discretization steps for ML-1M and Amazon-Beauty, and 50 for ML-20M. The number of sampling steps is fixed at 2. Further details are provided in Appendix~\ref{sec:implementation}.

\subsection{Performance Comparison}

\begin{table*}[t]
\caption{Performance comparison of multiple variants of \ours with different prior distributions and flows. The best results are highlighted in bold, and the second-best results are underlined.}
\label{tab:ablation}
\setlength{\tabcolsep}{3.5pt}

\begin{tabular}{cc|cccc|cccc|cccc}
\hline
& & \multicolumn{4}{c|}{\textbf{MovieLens-1M}} & \multicolumn{4}{c|}{\textbf{MovieLens-20M}} & \multicolumn{4}{c}{\textbf{Amazon-Beauty}} \\ \hline
\textbf{Methods} & \textbf{Priors} & \textbf{R@10} & \textbf{R@20}  & \textbf{N@10}  & \textbf{N@20}  & \textbf{R@10}  & \textbf{R@20}  & \textbf{N@10}  & \textbf{N@20}  & \textbf{R@10}  & \textbf{R@20}  & \textbf{N@10}  & \textbf{N@20}  \\ 
\hline
\textbf{C\ours} & \textbf{Uniform} & 0.2103 & 0.3048 & 0.2256 & 0.2457 & 0.2473 & 0.3437 & 0.2321 & 0.2546 & 0.1098 & 0.1447 & 0.0683 & 0.0775 \\
\textbf{C\ours} & \textbf{Gaussian} & 0.2116 & 0.3056 & 0.2276 & 0.2471 & 0.2523 & 0.3514 & 0.2341 & 0.2574 & 0.1189 & 0.1597 & 0.0728 & 0.0836 \\
\textbf{C\ours} & \textbf{Behavior-Guided} & \underline{0.2132} & \underline{0.3107} & \underline{0.2287} & \underline{0.2491} & \underline{0.2549} & \underline{0.3547} & \underline{0.2375} & \underline{0.2611} & \underline{0.1202} & \underline{0.1619} & \underline{0.0751} & \underline{0.0859} \\
\hline
\textbf{\ours} & \textbf{Random Binary} & 0.2035 & 0.2933 & 0.2214 & 0.2393 & 0.1888 & 0.2656 & 0.1724 & 0.1912 & 0.0018 & 0.0033 & 0.0009 & 0.0013 \\
\textbf{\ours} & \textbf{Uniform} & 0.2054 & 0.2942 & 0.2219 & 0.2394 & 0.0938 & 0.1511 & 0.0871 & 0.1021 & 0.0020 & 0.0032 & 0.0009 & 0.0012 \\
\textbf{\ours} & \textbf{Gaussian} & 0.2028 & 0.2937 & 0.2208 & 0.2392 & 0.1963 & 0.2699 & 0.1865 & 0.2039 & 0.0020 & 0.0031 & 0.0009 & 0.0011 \\
\textbf{\ours} & \textbf{Behavior-Guided} & \textbf{0.2157} & \textbf{0.3145} & \textbf{0.2320} & \textbf{0.2530} & \textbf{0.2580} & \textbf{0.3585} & \textbf{0.2394} & \textbf{0.2634} & \textbf{0.1245} & \textbf{0.1633} & \textbf{0.0755} & \textbf{0.0856} \\
\hline

\end{tabular}
\end{table*}

Table~\ref{tab:overall} presents the overall performance comparison of \ours and the baselines across three datasets. The results highlight several key observations and insights for generative CF and our \ours:
\begin{itemize}
[leftmargin=*,noitemsep,topsep=2pt]
	\item The generative CF methods demonstrate competitive performance compared to non-generative baseline methods. This suggests that generative CF models are promising for collaborative filtering, as they can effectively capture complex user-item interactions and generate high-quality recommendations by modeling the underlying interaction distribution.
	\item When compared to traditional generative models such as DAE and VAE-based methods, we observe that more advanced generative models, particularly the diffusion-based DiffRec and our proposed flow matching model, \ours, tend to achieve superior recommendation performance. This improvement can be attributed to the inherent strengths of diffusion models and flow matching techniques, which effectively capture intricate user behavior patterns and preference distributions, leading to more accurate and personalized recommendations.
	\item Our proposed \ours consistently outperforms all baseline models across all datasets and metrics with notable improvements. This demonstrates that \ours effectively combines the strengths of flow matching with the specific requirements of recommendation tasks. By incorporating our behavior-guided prior and discrete flow design for collaborative filtering, \ours is able to model user preference distribution more accurately than VAE and diffusion-based baseline models.
\end{itemize}

\begin{figure}[ht]
    \setlength{\abovecaptionskip}{-0.0cm}
    \setlength{\belowcaptionskip}{-0.3cm}
    \centering
    \includegraphics[width=0.47\textwidth]{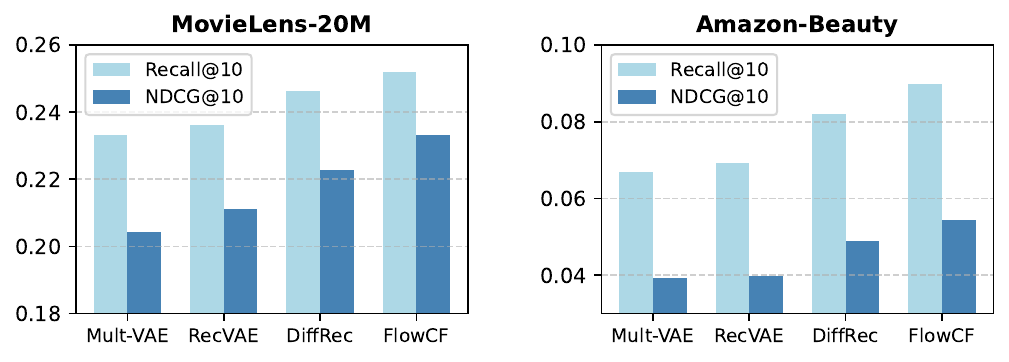}
    \caption{Performance comparison under natural noise setting on MovieLens-20M and Amazon-Beauty.}
    \label{fig:noise}
\end{figure}

\begin{figure}[ht]
	\setlength{\abovecaptionskip}{0.1cm}
    \setlength{\belowcaptionskip}{0.2cm}
    \includegraphics[width=0.45\textwidth]{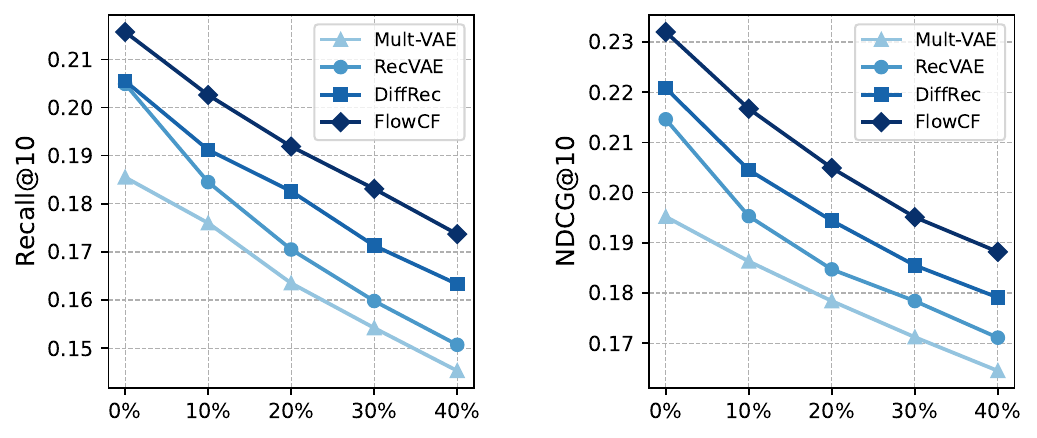}
    \caption{Performance of noisy training with different proportions of random noise on MovieLens-1M.}
    \label{fig:random_noise}
\end{figure}

\noindent \textbf{Noisy Training.} In real-world scenarios, implicit feedback often contains false-positive interactions.
To evaluate the robustness of \ours to such natural noise, we conduct experiments by randomly adding ratings below 4 as positive implicit feedback in training and validation sets. The results, illustrated in Figure~\ref{fig:noise}, reveal that our approach not only consistently outperforms baseline models under conditions of natural noise but also exhibits a lower rate of performance degradation compared to VAE-based methods. Furthermore, Figure~\ref{fig:random_noise} presents the experimental results involving random noise, with noise proportions ranging from 10\% to 50\% on the MovieLens-1M dataset. These results indicate that \ours maintains superior performance over both VAE and diffusion-based baseline models, demonstrating its robustness to random noise.
This robustness can be attributed to our training strategy, which leverages a set of interpolations that incorporate varying levels of noisy interactions between samples from prior and historical interactions. This approach enhances generalization by enabling the model to effectively transform between prior and historical interactions through carefully designed probability paths. These paths inherently accommodate noisy data, allowing \ours to learn more stable transformations and adapt effectively to noisy real-world conditions.

\vspace{-0.3cm}

\subsection{Ablation and Sensitivity Study}

\noindent \textbf{Effect of Priors and Flows.}
To study the effect of different prior distributions on both continuous flow-based recommender (denoted as C\ours, further details in Appendix~\ref{sec:cflowcf}) and discrete flow-based \ours, we compare uniform distribution between 0 and 1, Gaussian distribution $\mathcal N (0, I)$, and our behavior-guided prior distribution. For discrete flow-based \ours, we discretize these priors to probability mass functions using a Bernoulli distribution and also include a random binary initialization (0 or 1) as a baseline.

Table~\ref{tab:ablation} presents the performance of different variants with various priors and flows. The behavior-guided prior consistently outperforms the other priors, indicating that incorporating user behavior information into the prior facilitates the training of \ours. This validates our previous hypothesis that a behavior-guided prior, which better approximates the true user behavior distribution, enhances the performance of flow matching in collaborative filtering. While C\ours with uniform and Gaussian priors still achieves decent performance, the discrete flow-based \ours relies more heavily on the behavior-guided prior for stable training. Notably, using random binary initialization, discretized uniform distribution, or discretized Gaussian distribution as the prior can lead to training failures, as evidenced in the Amazon-Beauty dataset. 

These results indicate that although the discrete flow framework improves recommendation performance, it is more sensitive to the choice of prior. This sensitivity arises from the inherent discrete nature and the density of interaction data.
In discrete flow framework, the sample $X_0$ from prior is binary (only 0s and 1s). To learn effective transformations, the density of positives (1s) in $X_0$ should closely match the true density of the dataset. Dense samples from uniform or Gaussian priors introduce many false positives, hindering the model to recover correct patterns. In contrast, C\ours uses non-binary continuous-valued samples, which are not affected by the density of positives in discrete modeling. This explains its lower sensitivity to prior design compared to \ours.
Consequently, a customized behavior-guided prior for effective training. Additionally, we observe that the convergence speed is significantly slower when using priors other than the behavior-guided prior for \ours. Overall, these results highlight the importance of incorporating user behavior patterns into the prior distribution to effectively model implicit feedback and enhance recommendation performance in flow matching recommenders.

\begin{figure}[ht]
	\setlength{\abovecaptionskip}{-0.0cm}
    \setlength{\belowcaptionskip}{-0.2cm}
    \includegraphics[width=0.48\textwidth]{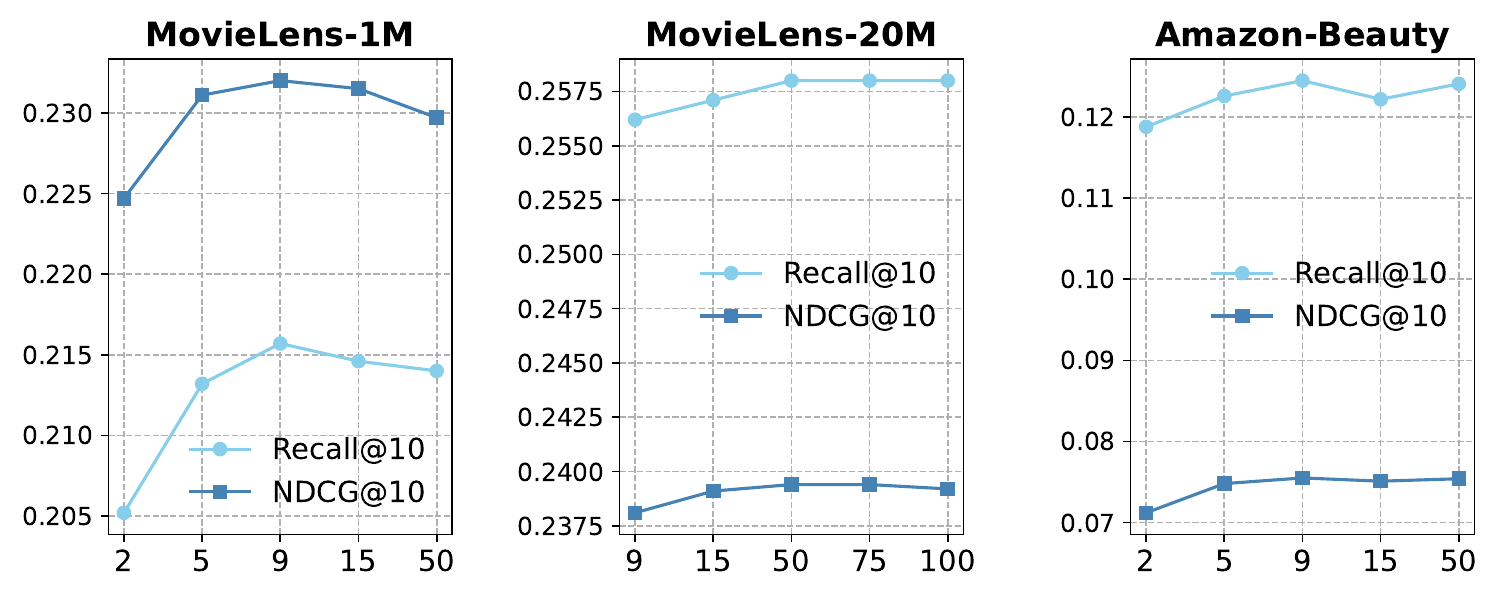}
    \caption{Effect of the number of discretization steps $N$.}
    \label{fig:step}
\end{figure}

\vspace{3pt}
\noindent \textbf{Sensitivity to Discretization Step.}
The number of discretization steps $N$ determines the granularity of the interpolation set. To study its impact, we conduct experiments with different values of $N$. The results in Figure~\ref{fig:step} indicate that the optimal choice of $N$ depends on the dataset size and its inherent sparsity. For example, MovieLens-1M and Amazon-Beauty achieve the peak results with $N = 9$, while MovieLens-20M performs best with $N$ between 50 and 75. Larger datasets generally benefit from finer discretization. Unlike diffusion-based models that require tuning noise schedules, our approach has fewer sensitive hyperparameters, with the primary focus being the selection of $N$ for interpolation.

\subsection{Efficiency Analysis}
To evaluate the efficiency of \ours, we compare its performance with VAE-based models and diffusion-based DiffRec in terms of training speed, inference speed, and convergence rate using the MovieLens-20M dataset, which contains the highest number of interactions among the datasets. All experiments are conducted on a single NVIDIA A5000 GPU under a fair baseline setting.

\begin{figure}[ht]
	\setlength{\abovecaptionskip}{0.1cm}
    \setlength{\belowcaptionskip}{-0.4cm}
    \centering
    \includegraphics[width=0.48\textwidth]{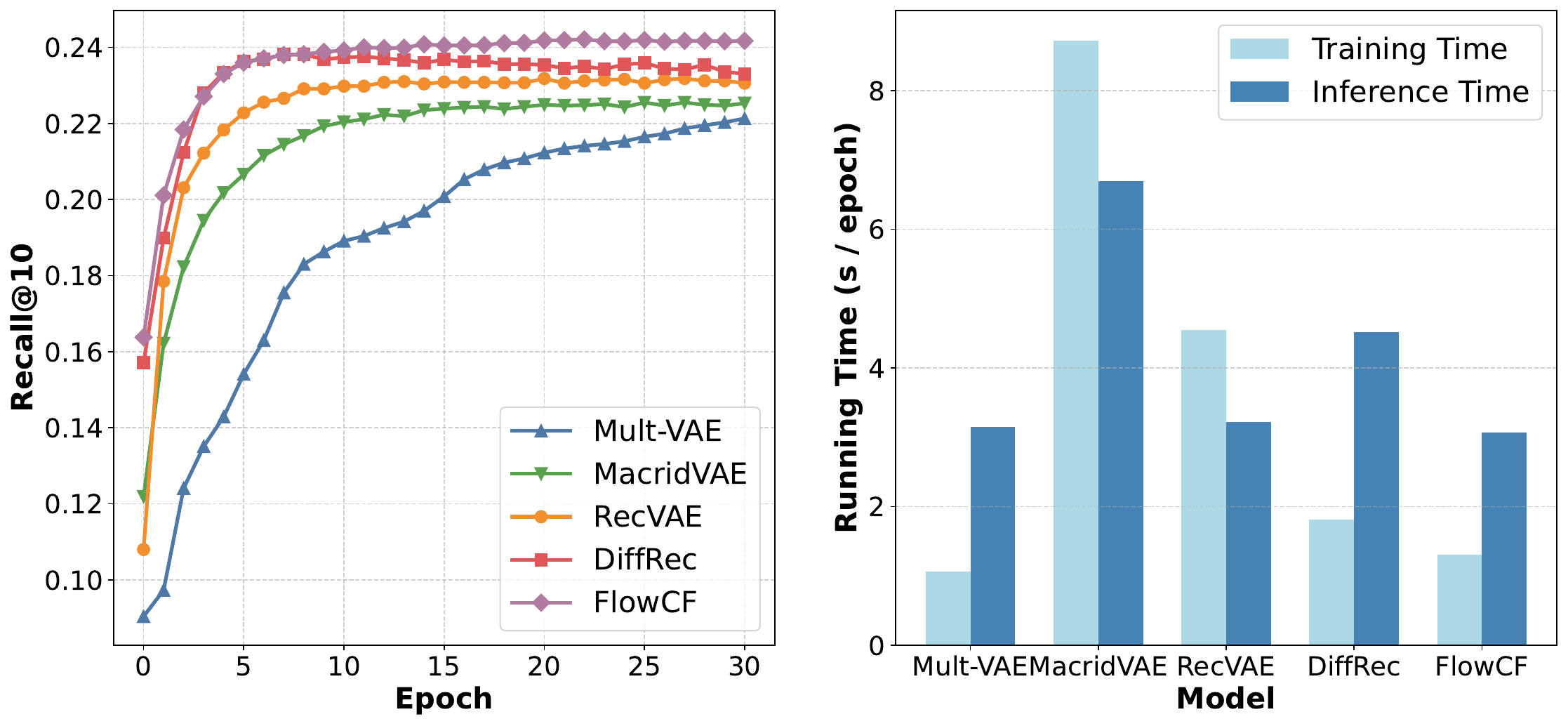}
    \caption{Learning curves and running time per epoch comparison of different generative CF models on ML-20M. The proposed \ours converges significantly faster while achieving superior performance and efficient inference speed.}
    \label{fig:efficiency}
\end{figure}

The learning curves on the left side of Figure~\ref{fig:efficiency} illustrate that among generative collaborative filtering models, \ours and DiffRec achieve the fastest convergence, whereas VAE-based methods converge more slowly. Specifically, Mult-VAE demonstrates fast per-epoch training speed but suffers from a slow convergence rate, which limits its overall training efficiency. RecVAE addresses this limitation by improving the convergence rate but at the cost of increased training time per epoch due to architectural modifications introduced to enhance Mult-VAE. In contrast, \ours combines rapid per-epoch training with a fast convergence rate, showcasing its superiority in balancing these two critical aspects of model efficiency.
The inference time per epoch, depicted on the right side of Figure~\ref{fig:efficiency}, further highlights the advantages of \ours. It achieves the fastest inference speed comparable to Mult-VAE and RecVAE by leveraging efficient sampling with only two steps during inference. Compared to DiffRec, which requires a larger number of sampling steps during inference due to its diffusion process, \ours significantly reduces this requirement.
Overall, \ours achieves both efficient training with rapid convergence and efficient inference, making it a highly competitive choice for recommendation systems.

\vspace{-3pt}

\section{Related Work}

\subsection{Collaborative Filtering}

\noindent \textbf{Non-Generative Collaborative Filtering.}
Non-generative collaborative filtering learns user and item representations in a discriminative manner. Most non-generative CF models rely on Bayesian Personalized Ranking loss~\cite{rendle2012bpr} along with the application of deep learning techniques~\cite{he2017neural}, such as graph neural networks~\cite{wang2019neural, he2020lightgcn}, contrastive learning~\cite{wujc2021self, lin2022improving, yu2022graph, liu2024twincl}, and diffusion models~\cite{zhao2024denoising, yi2024directional} to enhance more robust and informative representation learning.

\noindent \textbf{Generative Collaborative Filtering.}
The generative collaborative filtering methods take a different approach from learning user and item representations. These methods aim to directly reconstruct user-item interaction matrices by generating probability distributions. The denoising autoencoders (DAEs)~\cite{vincent2008extracting, bengio2013generalized} and variational autoencoders (VAEs)~\cite{kingma2013auto} have been prevalent in generative CF. The DAE-based CF models~\cite{wu2016collaborative, liang2018variational} encode corrupted user-item interactions into latent representations and decode them to reconstruct the original input, learning to denoise and capture underlying patterns. The VAE-based recommenders~\cite{liang2018variational, ma2019learning, shenbin2020recvae} encode user-item interactions into latent probability distributions and decode these distributions to predict user preferences and interactions.
Recently, diffusion models~\cite{lin2024survey} have also shown promise in generative CF. DiffRec~\cite{wang2023diffusion} is the pioneer in using diffusion models in collaborative filtering by modeling the distributions of user-item interaction probabilities. GiffCF~\cite{zhu2024graph} and CF-Diff~\cite{hou2024collaborative} further advance and enhance diffusion-based CF models by incorporating graph signal processing and attention mechanisms, respectively.

\vspace{-3pt}
\subsection{Generative Models}

\noindent \textbf{Diffusion Models}  
~\cite{sohl2015deep, song2019generative, ho2020denoising, song2020score} generate data by iteratively refining samples from Gaussian noise to meaningful outputs. These models have achieved strong results across domains, including images~\cite{ho2020denoising, rombach2022high}, molecules~\cite{xu2022geodiff, jing2022torsional}, material science~\cite{xie2021crystal, luo2023towards} and recommendation systems~\cite{lin2024survey, li2023diffurec, yang2024generate}. 

\noindent \textbf{Flow Matching}
~\cite{lipman2022flow, liu2022flow, albergo2022building} is a flow-based method using continuous normalizing flows~\cite{chen2018neural} as alternatives to diffusion models. Conditional Flow Matching~\cite{lipman2022flow, tong2023improving} directly learns a vector field that defines the probability flow from a source distribution to the target distribution, conditioned on individual data points. Additionally, recent advances have also extended the flow matching framework to discrete state spaces~\cite{campbell2024generative, gat2024discrete}. Flow matching offers faster inference and improved efficiency and has been applied to various generative tasks such as images~\cite{esser2403scaling}, videos~\cite{polyak2024movie}, and proteins~\cite{bose2023se}. However, its potential in recommender systems remains unexplored, presenting opportunities for future research.  

\vspace{-2pt}
\section{Conclusion}
In this paper, we introduce \ours, a novel flow-based recommender system that utilizes flow matching for generative collaborative filtering. To adapt flow matching to recommendation tasks, we propose a behavior-guided prior distribution to guide the flow training process with user behavior patterns, and a discrete flow framework that effectively models discrete and binary implicit feedback. Comprehensive experiments demonstrate that our \ours achieves state-of-the-art recommendation performance with efficient inference speed. 
The success of \ours opens promising directions for future research on flow matching in recommender systems and highlights its potential for advancing the field of generative collaborative filtering.

\begin{acks}
We thank the anonymous reviewers and area chairs for their valuable feedback and suggestions.
This research was supported in part by Yale AI Engineering Research Seed Grants, Yale Office of the Provost, and Amazon Research Award to the Yale University team.
\end{acks}

\bibliographystyle{ACM-Reference-Format}
\balance

\bibliography{ref}


\begin{thebibliography}{53}


\ifx \showCODEN    \undefined \def \showCODEN     #1{\unskip}     \fi
\ifx \showDOI      \undefined \def \showDOI       #1{#1}\fi
\ifx \showISBNx    \undefined \def \showISBNx     #1{\unskip}     \fi
\ifx \showISBNxiii \undefined \def \showISBNxiii  #1{\unskip}     \fi
\ifx \showISSN     \undefined \def \showISSN      #1{\unskip}     \fi
\ifx \showLCCN     \undefined \def \showLCCN      #1{\unskip}     \fi
\ifx \shownote     \undefined \def \shownote      #1{#1}          \fi
\ifx \showarticletitle \undefined \def \showarticletitle #1{#1}   \fi
\ifx \showURL      \undefined \def \showURL       {\relax}        \fi
\providecommand\bibfield[2]{#2}
\providecommand\bibinfo[2]{#2}
\providecommand\natexlab[1]{#1}
\providecommand\showeprint[2][]{arXiv:#2}

\bibitem[Albergo and Vanden-Eijnden(2022)]%
        {albergo2022building}
\bibfield{author}{\bibinfo{person}{Michael~S Albergo} {and} \bibinfo{person}{Eric Vanden-Eijnden}.} \bibinfo{year}{2022}\natexlab{}.
\newblock \showarticletitle{Building normalizing flows with stochastic interpolants}.
\newblock \bibinfo{journal}{\emph{arXiv preprint arXiv:2209.15571}} (\bibinfo{year}{2022}).
\newblock


\bibitem[Bengio et~al\mbox{.}(2013)]%
        {bengio2013generalized}
\bibfield{author}{\bibinfo{person}{Yoshua Bengio}, \bibinfo{person}{Li Yao}, \bibinfo{person}{Guillaume Alain}, {and} \bibinfo{person}{Pascal Vincent}.} \bibinfo{year}{2013}\natexlab{}.
\newblock \showarticletitle{Generalized denoising auto-encoders as generative models}.
\newblock \bibinfo{journal}{\emph{Advances in neural information processing systems}}  \bibinfo{volume}{26} (\bibinfo{year}{2013}).
\newblock


\bibitem[Bose et~al\mbox{.}(2023)]%
        {bose2023se}
\bibfield{author}{\bibinfo{person}{Avishek~Joey Bose}, \bibinfo{person}{Tara Akhound-Sadegh}, \bibinfo{person}{Guillaume Huguet}, \bibinfo{person}{Kilian Fatras}, \bibinfo{person}{Jarrid Rector-Brooks}, \bibinfo{person}{Cheng-Hao Liu}, \bibinfo{person}{Andrei~Cristian Nica}, \bibinfo{person}{Maksym Korablyov}, \bibinfo{person}{Michael Bronstein}, {and} \bibinfo{person}{Alexander Tong}.} \bibinfo{year}{2023}\natexlab{}.
\newblock \showarticletitle{Se (3)-stochastic flow matching for protein backbone generation}.
\newblock \bibinfo{journal}{\emph{arXiv preprint arXiv:2310.02391}} (\bibinfo{year}{2023}).
\newblock


\bibitem[Campbell et~al\mbox{.}(2024)]%
        {campbell2024generative}
\bibfield{author}{\bibinfo{person}{Andrew Campbell}, \bibinfo{person}{Jason Yim}, \bibinfo{person}{Regina Barzilay}, \bibinfo{person}{Tom Rainforth}, {and} \bibinfo{person}{Tommi Jaakkola}.} \bibinfo{year}{2024}\natexlab{}.
\newblock \showarticletitle{Generative flows on discrete state-spaces: Enabling multimodal flows with applications to protein co-design}.
\newblock \bibinfo{journal}{\emph{arXiv preprint arXiv:2402.04997}} (\bibinfo{year}{2024}).
\newblock


\bibitem[Chen et~al\mbox{.}(2022)]%
        {chen2022generative}
\bibfield{author}{\bibinfo{person}{Hao Chen}, \bibinfo{person}{Zefan Wang}, \bibinfo{person}{Feiran Huang}, \bibinfo{person}{Xiao Huang}, \bibinfo{person}{Yue Xu}, \bibinfo{person}{Yishi Lin}, \bibinfo{person}{Peng He}, {and} \bibinfo{person}{Zhoujun Li}.} \bibinfo{year}{2022}\natexlab{}.
\newblock \showarticletitle{Generative adversarial framework for cold-start item recommendation}. In \bibinfo{booktitle}{\emph{Proceedings of the 45th International ACM SIGIR Conference on Research and Development in Information Retrieval}}. \bibinfo{pages}{2565--2571}.
\newblock


\bibitem[Chen et~al\mbox{.}(2018)]%
        {chen2018neural}
\bibfield{author}{\bibinfo{person}{Ricky~TQ Chen}, \bibinfo{person}{Yulia Rubanova}, \bibinfo{person}{Jesse Bettencourt}, {and} \bibinfo{person}{David~K Duvenaud}.} \bibinfo{year}{2018}\natexlab{}.
\newblock \showarticletitle{Neural ordinary differential equations}.
\newblock \bibinfo{journal}{\emph{Advances in neural information processing systems}}  \bibinfo{volume}{31} (\bibinfo{year}{2018}).
\newblock


\bibitem[Esser et~al\mbox{.}({[n.\,d.]})]%
        {esser2403scaling}
\bibfield{author}{\bibinfo{person}{Patrick Esser}, \bibinfo{person}{Sumith Kulal}, \bibinfo{person}{Andreas Blattmann}, \bibinfo{person}{Rahim Entezari}, \bibinfo{person}{Jonas M{\"u}ller}, \bibinfo{person}{Harry Saini}, \bibinfo{person}{Yam Levi}, \bibinfo{person}{Dominik Lorenz}, \bibinfo{person}{Axel Sauer}, \bibinfo{person}{Frederic Boesel}, {et~al\mbox{.}}} \bibinfo{year}{[n.\,d.]}\natexlab{}.
\newblock \showarticletitle{Scaling rectified flow transformers for high-resolution image synthesis, 2024}.
\newblock \bibinfo{journal}{\emph{URL https://arxiv. org/abs/2403.03206}}  \bibinfo{volume}{2} (\bibinfo{year}{[n.\,d.]}).
\newblock


\bibitem[Gat et~al\mbox{.}(2024)]%
        {gat2024discrete}
\bibfield{author}{\bibinfo{person}{Itai Gat}, \bibinfo{person}{Tal Remez}, \bibinfo{person}{Neta Shaul}, \bibinfo{person}{Felix Kreuk}, \bibinfo{person}{Ricky~TQ Chen}, \bibinfo{person}{Gabriel Synnaeve}, \bibinfo{person}{Yossi Adi}, {and} \bibinfo{person}{Yaron Lipman}.} \bibinfo{year}{2024}\natexlab{}.
\newblock \showarticletitle{Discrete flow matching}.
\newblock \bibinfo{journal}{\emph{arXiv preprint arXiv:2407.15595}} (\bibinfo{year}{2024}).
\newblock


\bibitem[Harper and Konstan(2015)]%
        {harper2015movielens}
\bibfield{author}{\bibinfo{person}{F~Maxwell Harper} {and} \bibinfo{person}{Joseph~A Konstan}.} \bibinfo{year}{2015}\natexlab{}.
\newblock \showarticletitle{The movielens datasets: History and context}.
\newblock \bibinfo{journal}{\emph{Acm transactions on interactive intelligent systems (tiis)}} (\bibinfo{year}{2015}).
\newblock


\bibitem[He et~al\mbox{.}(2020)]%
        {he2020lightgcn}
\bibfield{author}{\bibinfo{person}{Xiangnan He}, \bibinfo{person}{Kuan Deng}, \bibinfo{person}{Xiang Wang}, \bibinfo{person}{Yan Li}, \bibinfo{person}{Yongdong Zhang}, {and} \bibinfo{person}{Meng Wang}.} \bibinfo{year}{2020}\natexlab{}.
\newblock \showarticletitle{Lightgcn: Simplifying and powering graph convolution network for recommendation}. In \bibinfo{booktitle}{\emph{Proceedings of the 43rd International ACM SIGIR conference on research and development in Information Retrieval}}. \bibinfo{pages}{639--648}.
\newblock


\bibitem[He et~al\mbox{.}(2017)]%
        {he2017neural}
\bibfield{author}{\bibinfo{person}{Xiangnan He}, \bibinfo{person}{Lizi Liao}, \bibinfo{person}{Hanwang Zhang}, \bibinfo{person}{Liqiang Nie}, \bibinfo{person}{Xia Hu}, {and} \bibinfo{person}{Tat-Seng Chua}.} \bibinfo{year}{2017}\natexlab{}.
\newblock \showarticletitle{Neural collaborative filtering}. In \bibinfo{booktitle}{\emph{Proceedings of the 26th international conference on world wide web}}. \bibinfo{pages}{173--182}.
\newblock


\bibitem[Ho et~al\mbox{.}(2020)]%
        {ho2020denoising}
\bibfield{author}{\bibinfo{person}{Jonathan Ho}, \bibinfo{person}{Ajay Jain}, {and} \bibinfo{person}{Pieter Abbeel}.} \bibinfo{year}{2020}\natexlab{}.
\newblock \showarticletitle{Denoising diffusion probabilistic models}.
\newblock \bibinfo{journal}{\emph{Advances in neural information processing systems}}  \bibinfo{volume}{33} (\bibinfo{year}{2020}), \bibinfo{pages}{6840--6851}.
\newblock


\bibitem[Hou et~al\mbox{.}(2024)]%
        {hou2024collaborative}
\bibfield{author}{\bibinfo{person}{Yu Hou}, \bibinfo{person}{Jin-Duk Park}, {and} \bibinfo{person}{Won-Yong Shin}.} \bibinfo{year}{2024}\natexlab{}.
\newblock \showarticletitle{Collaborative filtering based on diffusion models: Unveiling the potential of high-order connectivity}. In \bibinfo{booktitle}{\emph{Proceedings of the 47th International ACM SIGIR Conference on Research and Development in Information Retrieval}}. \bibinfo{pages}{1360--1369}.
\newblock


\bibitem[Jing et~al\mbox{.}(2022)]%
        {jing2022torsional}
\bibfield{author}{\bibinfo{person}{Bowen Jing}, \bibinfo{person}{Gabriele Corso}, \bibinfo{person}{Jeffrey Chang}, \bibinfo{person}{Regina Barzilay}, {and} \bibinfo{person}{Tommi Jaakkola}.} \bibinfo{year}{2022}\natexlab{}.
\newblock \showarticletitle{Torsional diffusion for molecular conformer generation}.
\newblock \bibinfo{journal}{\emph{Advances in Neural Information Processing Systems}}  \bibinfo{volume}{35} (\bibinfo{year}{2022}), \bibinfo{pages}{24240--24253}.
\newblock


\bibitem[Kingma(2013)]%
        {kingma2013auto}
\bibfield{author}{\bibinfo{person}{Diederik~P Kingma}.} \bibinfo{year}{2013}\natexlab{}.
\newblock \showarticletitle{Auto-encoding variational bayes}.
\newblock \bibinfo{journal}{\emph{arXiv preprint arXiv:1312.6114}} (\bibinfo{year}{2013}).
\newblock


\bibitem[Kingma(2014)]%
        {kingma2014adam}
\bibfield{author}{\bibinfo{person}{Diederik~P Kingma}.} \bibinfo{year}{2014}\natexlab{}.
\newblock \showarticletitle{Adam: A method for stochastic optimization}.
\newblock \bibinfo{journal}{\emph{arXiv preprint arXiv:1412.6980}} (\bibinfo{year}{2014}).
\newblock


\bibitem[Koren et~al\mbox{.}(2021)]%
        {koren2021advances}
\bibfield{author}{\bibinfo{person}{Yehuda Koren}, \bibinfo{person}{Steffen Rendle}, {and} \bibinfo{person}{Robert Bell}.} \bibinfo{year}{2021}\natexlab{}.
\newblock \showarticletitle{Advances in collaborative filtering}.
\newblock \bibinfo{journal}{\emph{Recommender systems handbook}} (\bibinfo{year}{2021}), \bibinfo{pages}{91--142}.
\newblock


\bibitem[Li et~al\mbox{.}(2023)]%
        {li2023diffurec}
\bibfield{author}{\bibinfo{person}{Zihao Li}, \bibinfo{person}{Aixin Sun}, {and} \bibinfo{person}{Chenliang Li}.} \bibinfo{year}{2023}\natexlab{}.
\newblock \showarticletitle{Diffurec: A diffusion model for sequential recommendation}.
\newblock \bibinfo{journal}{\emph{ACM Transactions on Information Systems}} \bibinfo{volume}{42}, \bibinfo{number}{3} (\bibinfo{year}{2023}), \bibinfo{pages}{1--28}.
\newblock


\bibitem[Liang et~al\mbox{.}(2018)]%
        {liang2018variational}
\bibfield{author}{\bibinfo{person}{Dawen Liang}, \bibinfo{person}{Rahul~G Krishnan}, \bibinfo{person}{Matthew~D Hoffman}, {and} \bibinfo{person}{Tony Jebara}.} \bibinfo{year}{2018}\natexlab{}.
\newblock \showarticletitle{Variational autoencoders for collaborative filtering}. In \bibinfo{booktitle}{\emph{Proceedings of the 2018 world wide web conference}}. \bibinfo{pages}{689--698}.
\newblock


\bibitem[Lin et~al\mbox{.}(2024)]%
        {lin2024survey}
\bibfield{author}{\bibinfo{person}{Jianghao Lin}, \bibinfo{person}{Jiaqi Liu}, \bibinfo{person}{Jiachen Zhu}, \bibinfo{person}{Yunjia Xi}, \bibinfo{person}{Chengkai Liu}, \bibinfo{person}{Yangtian Zhang}, \bibinfo{person}{Yong Yu}, {and} \bibinfo{person}{Weinan Zhang}.} \bibinfo{year}{2024}\natexlab{}.
\newblock \showarticletitle{A Survey on Diffusion Models for Recommender Systems}.
\newblock \bibinfo{journal}{\emph{arXiv preprint arXiv:2409.05033}} (\bibinfo{year}{2024}).
\newblock


\bibitem[Lin et~al\mbox{.}(2022)]%
        {lin2022improving}
\bibfield{author}{\bibinfo{person}{Zihan Lin}, \bibinfo{person}{Changxin Tian}, \bibinfo{person}{Yupeng Hou}, {and} \bibinfo{person}{Wayne~Xin Zhao}.} \bibinfo{year}{2022}\natexlab{}.
\newblock \showarticletitle{Improving graph collaborative filtering with neighborhood-enriched contrastive learning}. In \bibinfo{booktitle}{\emph{Proceedings of the ACM web conference 2022}}. \bibinfo{pages}{2320--2329}.
\newblock


\bibitem[Lipman et~al\mbox{.}(2022)]%
        {lipman2022flow}
\bibfield{author}{\bibinfo{person}{Yaron Lipman}, \bibinfo{person}{Ricky~TQ Chen}, \bibinfo{person}{Heli Ben-Hamu}, \bibinfo{person}{Maximilian Nickel}, {and} \bibinfo{person}{Matt Le}.} \bibinfo{year}{2022}\natexlab{}.
\newblock \showarticletitle{Flow matching for generative modeling}.
\newblock \bibinfo{journal}{\emph{arXiv preprint arXiv:2210.02747}} (\bibinfo{year}{2022}).
\newblock


\bibitem[Liu et~al\mbox{.}(2024a)]%
        {liu2024behavior}
\bibfield{author}{\bibinfo{person}{Chengkai Liu}, \bibinfo{person}{Jianghao Lin}, \bibinfo{person}{Hanzhou Liu}, \bibinfo{person}{Jianling Wang}, {and} \bibinfo{person}{James Caverlee}.} \bibinfo{year}{2024}\natexlab{a}.
\newblock \showarticletitle{Behavior-Dependent Linear Recurrent Units for Efficient Sequential Recommendation}.
\newblock \bibinfo{journal}{\emph{arXiv preprint arXiv:2406.12580}} (\bibinfo{year}{2024}).
\newblock


\bibitem[Liu et~al\mbox{.}(2024b)]%
        {liu2024mamba4rec}
\bibfield{author}{\bibinfo{person}{Chengkai Liu}, \bibinfo{person}{Jianghao Lin}, \bibinfo{person}{Jianling Wang}, \bibinfo{person}{Hanzhou Liu}, {and} \bibinfo{person}{James Caverlee}.} \bibinfo{year}{2024}\natexlab{b}.
\newblock \showarticletitle{Mamba4rec: Towards efficient sequential recommendation with selective state space models}.
\newblock \bibinfo{journal}{\emph{arXiv preprint arXiv:2403.03900}} (\bibinfo{year}{2024}).
\newblock


\bibitem[Liu et~al\mbox{.}(2024c)]%
        {liu2024twincl}
\bibfield{author}{\bibinfo{person}{Chengkai Liu}, \bibinfo{person}{Jianling Wang}, {and} \bibinfo{person}{James Caverlee}.} \bibinfo{year}{2024}\natexlab{c}.
\newblock \showarticletitle{TwinCL: A Twin Graph Contrastive Learning Model for Collaborative Filtering}.
\newblock \bibinfo{journal}{\emph{arXiv preprint arXiv:2409.19169}} (\bibinfo{year}{2024}).
\newblock


\bibitem[Liu et~al\mbox{.}(2022)]%
        {liu2022flow}
\bibfield{author}{\bibinfo{person}{Xingchao Liu}, \bibinfo{person}{Chengyue Gong}, {and} \bibinfo{person}{Qiang Liu}.} \bibinfo{year}{2022}\natexlab{}.
\newblock \showarticletitle{Flow straight and fast: Learning to generate and transfer data with rectified flow}.
\newblock \bibinfo{journal}{\emph{arXiv preprint arXiv:2209.03003}} (\bibinfo{year}{2022}).
\newblock


\bibitem[Luo et~al\mbox{.}(2023)]%
        {luo2023towards}
\bibfield{author}{\bibinfo{person}{Youzhi Luo}, \bibinfo{person}{Chengkai Liu}, {and} \bibinfo{person}{Shuiwang Ji}.} \bibinfo{year}{2023}\natexlab{}.
\newblock \showarticletitle{Towards symmetry-aware generation of periodic materials}.
\newblock \bibinfo{journal}{\emph{Advances in Neural Information Processing Systems}}  \bibinfo{volume}{36} (\bibinfo{year}{2023}), \bibinfo{pages}{53308--53329}.
\newblock


\bibitem[Ma et~al\mbox{.}(2019)]%
        {ma2019learning}
\bibfield{author}{\bibinfo{person}{Jianxin Ma}, \bibinfo{person}{Chang Zhou}, \bibinfo{person}{Peng Cui}, \bibinfo{person}{Hongxia Yang}, {and} \bibinfo{person}{Wenwu Zhu}.} \bibinfo{year}{2019}\natexlab{}.
\newblock \showarticletitle{Learning disentangled representations for recommendation}.
\newblock \bibinfo{journal}{\emph{Advances in neural information processing systems}}  \bibinfo{volume}{32} (\bibinfo{year}{2019}).
\newblock


\bibitem[McAuley et~al\mbox{.}(2015)]%
        {mcauley2015image}
\bibfield{author}{\bibinfo{person}{Julian McAuley}, \bibinfo{person}{Christopher Targett}, \bibinfo{person}{Qinfeng Shi}, {and} \bibinfo{person}{Anton Van Den~Hengel}.} \bibinfo{year}{2015}\natexlab{}.
\newblock \showarticletitle{Image-based recommendations on styles and substitutes}. In \bibinfo{booktitle}{\emph{Proceedings of the 38th international ACM SIGIR conference on research and development in information retrieval}}. \bibinfo{pages}{43--52}.
\newblock


\bibitem[Papamakarios et~al\mbox{.}(2021)]%
        {papamakarios2021normalizing}
\bibfield{author}{\bibinfo{person}{George Papamakarios}, \bibinfo{person}{Eric Nalisnick}, \bibinfo{person}{Danilo~Jimenez Rezende}, \bibinfo{person}{Shakir Mohamed}, {and} \bibinfo{person}{Balaji Lakshminarayanan}.} \bibinfo{year}{2021}\natexlab{}.
\newblock \showarticletitle{Normalizing flows for probabilistic modeling and inference}.
\newblock \bibinfo{journal}{\emph{Journal of Machine Learning Research}} \bibinfo{volume}{22}, \bibinfo{number}{57} (\bibinfo{year}{2021}), \bibinfo{pages}{1--64}.
\newblock


\bibitem[Polyak et~al\mbox{.}(2024)]%
        {polyak2024movie}
\bibfield{author}{\bibinfo{person}{Adam Polyak}, \bibinfo{person}{Amit Zohar}, \bibinfo{person}{Andrew Brown}, \bibinfo{person}{Andros Tjandra}, \bibinfo{person}{Animesh Sinha}, \bibinfo{person}{Ann Lee}, \bibinfo{person}{Apoorv Vyas}, \bibinfo{person}{Bowen Shi}, \bibinfo{person}{Chih-Yao Ma}, \bibinfo{person}{Ching-Yao Chuang}, {et~al\mbox{.}}} \bibinfo{year}{2024}\natexlab{}.
\newblock \showarticletitle{Movie gen: A cast of media foundation models}.
\newblock \bibinfo{journal}{\emph{arXiv preprint arXiv:2410.13720}}.
\newblock


\bibitem[Rendle et~al\mbox{.}(2012)]%
        {rendle2012bpr}
\bibfield{author}{\bibinfo{person}{Steffen Rendle}, \bibinfo{person}{Christoph Freudenthaler}, \bibinfo{person}{Zeno Gantner}, {and} \bibinfo{person}{Lars Schmidt-Thieme}.} \bibinfo{year}{2012}\natexlab{}.
\newblock \showarticletitle{BPR: Bayesian personalized ranking from implicit feedback}.
\newblock \bibinfo{journal}{\emph{arXiv preprint arXiv:1205.2618}} (\bibinfo{year}{2012}).
\newblock


\bibitem[Rombach et~al\mbox{.}(2022)]%
        {rombach2022high}
\bibfield{author}{\bibinfo{person}{Robin Rombach}, \bibinfo{person}{Andreas Blattmann}, \bibinfo{person}{Dominik Lorenz}, \bibinfo{person}{Patrick Esser}, {and} \bibinfo{person}{Bj{\"o}rn Ommer}.} \bibinfo{year}{2022}\natexlab{}.
\newblock \showarticletitle{High-resolution image synthesis with latent diffusion models}. In \bibinfo{booktitle}{\emph{Proceedings of the IEEE/CVF conference on computer vision and pattern recognition}}. \bibinfo{pages}{10684--10695}.
\newblock


\bibitem[Schafer et~al\mbox{.}(2007)]%
        {schafer2007collaborative}
\bibfield{author}{\bibinfo{person}{J~Ben Schafer}, \bibinfo{person}{Dan Frankowski}, \bibinfo{person}{Jon Herlocker}, {and} \bibinfo{person}{Shilad Sen}.} \bibinfo{year}{2007}\natexlab{}.
\newblock \showarticletitle{Collaborative filtering recommender systems}.
\newblock In \bibinfo{booktitle}{\emph{The adaptive web: methods and strategies of web personalization}}. \bibinfo{publisher}{Springer}, \bibinfo{pages}{291--324}.
\newblock


\bibitem[Shenbin et~al\mbox{.}(2020)]%
        {shenbin2020recvae}
\bibfield{author}{\bibinfo{person}{Ilya Shenbin}, \bibinfo{person}{Anton Alekseev}, \bibinfo{person}{Elena Tutubalina}, \bibinfo{person}{Valentin Malykh}, {and} \bibinfo{person}{Sergey~I Nikolenko}.} \bibinfo{year}{2020}\natexlab{}.
\newblock \showarticletitle{Recvae: A new variational autoencoder for top-n recommendations with implicit feedback}. In \bibinfo{booktitle}{\emph{Proceedings of the 13th international conference on web search and data mining}}. \bibinfo{pages}{528--536}.
\newblock


\bibitem[Sohl-Dickstein et~al\mbox{.}(2015)]%
        {sohl2015deep}
\bibfield{author}{\bibinfo{person}{Jascha Sohl-Dickstein}, \bibinfo{person}{Eric Weiss}, \bibinfo{person}{Niru Maheswaranathan}, {and} \bibinfo{person}{Surya Ganguli}.} \bibinfo{year}{2015}\natexlab{}.
\newblock \showarticletitle{Deep unsupervised learning using nonequilibrium thermodynamics}. In \bibinfo{booktitle}{\emph{International conference on machine learning}}. PMLR, \bibinfo{pages}{2256--2265}.
\newblock


\bibitem[Song and Ermon(2019)]%
        {song2019generative}
\bibfield{author}{\bibinfo{person}{Yang Song} {and} \bibinfo{person}{Stefano Ermon}.} \bibinfo{year}{2019}\natexlab{}.
\newblock \showarticletitle{Generative modeling by estimating gradients of the data distribution}.
\newblock \bibinfo{journal}{\emph{Advances in neural information processing systems}}.
\newblock


\bibitem[Song et~al\mbox{.}(2020)]%
        {song2020score}
\bibfield{author}{\bibinfo{person}{Yang Song}, \bibinfo{person}{Jascha Sohl-Dickstein}, \bibinfo{person}{Diederik~P Kingma}, \bibinfo{person}{Abhishek Kumar}, \bibinfo{person}{Stefano Ermon}, {and} \bibinfo{person}{Ben Poole}.} \bibinfo{year}{2020}\natexlab{}.
\newblock \showarticletitle{Score-based generative modeling through stochastic differential equations}.
\newblock \bibinfo{journal}{\emph{arXiv preprint arXiv:2011.13456}} (\bibinfo{year}{2020}).
\newblock


\bibitem[Tong et~al\mbox{.}(2023)]%
        {tong2023improving}
\bibfield{author}{\bibinfo{person}{Alexander Tong}, \bibinfo{person}{Kilian Fatras}, \bibinfo{person}{Nikolay Malkin}, \bibinfo{person}{Guillaume Huguet}, \bibinfo{person}{Yanlei Zhang}, \bibinfo{person}{Jarrid Rector-Brooks}, \bibinfo{person}{Guy Wolf}, {and} \bibinfo{person}{Yoshua Bengio}.} \bibinfo{year}{2023}\natexlab{}.
\newblock \showarticletitle{Improving and generalizing flow-based generative models with minibatch optimal transport}.
\newblock \bibinfo{journal}{\emph{arXiv preprint arXiv:2302.00482}} (\bibinfo{year}{2023}).
\newblock


\bibitem[Vincent et~al\mbox{.}(2008)]%
        {vincent2008extracting}
\bibfield{author}{\bibinfo{person}{Pascal Vincent}, \bibinfo{person}{Hugo Larochelle}, \bibinfo{person}{Yoshua Bengio}, {and} \bibinfo{person}{Pierre-Antoine Manzagol}.} \bibinfo{year}{2008}\natexlab{}.
\newblock \showarticletitle{Extracting and composing robust features with denoising autoencoders}. In \bibinfo{booktitle}{\emph{Proceedings of the 25th international conference on Machine learning}}. \bibinfo{pages}{1096--1103}.
\newblock


\bibitem[Wang et~al\mbox{.}(2021)]%
        {wang2021denoising}
\bibfield{author}{\bibinfo{person}{Wenjie Wang}, \bibinfo{person}{Fuli Feng}, \bibinfo{person}{Xiangnan He}, \bibinfo{person}{Liqiang Nie}, {and} \bibinfo{person}{Tat-Seng Chua}.} \bibinfo{year}{2021}\natexlab{}.
\newblock \showarticletitle{Denoising implicit feedback for recommendation}. In \bibinfo{booktitle}{\emph{Proceedings of the 14th ACM international conference on web search and data mining}}. \bibinfo{pages}{373--381}.
\newblock


\bibitem[Wang et~al\mbox{.}(2023)]%
        {wang2023diffusion}
\bibfield{author}{\bibinfo{person}{Wenjie Wang}, \bibinfo{person}{Yiyan Xu}, \bibinfo{person}{Fuli Feng}, \bibinfo{person}{Xinyu Lin}, \bibinfo{person}{Xiangnan He}, {and} \bibinfo{person}{Tat-Seng Chua}.} \bibinfo{year}{2023}\natexlab{}.
\newblock \showarticletitle{Diffusion recommender model}. In \bibinfo{booktitle}{\emph{Proceedings of the 46th International ACM SIGIR Conference on Research and Development in Information Retrieval}}. \bibinfo{pages}{832--841}.
\newblock


\bibitem[Wang et~al\mbox{.}(2019)]%
        {wang2019neural}
\bibfield{author}{\bibinfo{person}{Xiang Wang}, \bibinfo{person}{Xiangnan He}, \bibinfo{person}{Meng Wang}, \bibinfo{person}{Fuli Feng}, {and} \bibinfo{person}{Tat-Seng Chua}.} \bibinfo{year}{2019}\natexlab{}.
\newblock \showarticletitle{Neural graph collaborative filtering}. In \bibinfo{booktitle}{\emph{Proceedings of the 42nd international ACM SIGIR conference on Research and development in Information Retrieval}}. \bibinfo{pages}{165--174}.
\newblock


\bibitem[Wu et~al\mbox{.}(2021)]%
        {wujc2021self}
\bibfield{author}{\bibinfo{person}{Jiancan Wu}, \bibinfo{person}{Xiang Wang}, \bibinfo{person}{Fuli Feng}, \bibinfo{person}{Xiangnan He}, \bibinfo{person}{Liang Chen}, \bibinfo{person}{Jianxun Lian}, {and} \bibinfo{person}{Xing Xie}.} \bibinfo{year}{2021}\natexlab{}.
\newblock \showarticletitle{Self-supervised graph learning for recommendation}. In \bibinfo{booktitle}{\emph{Proceedings of the 44th International ACM SIGIR Conference on Research and Development in Information Retrieval}}. \bibinfo{pages}{726--735}.
\newblock


\bibitem[Wu et~al\mbox{.}(2016)]%
        {wu2016collaborative}
\bibfield{author}{\bibinfo{person}{Yao Wu}, \bibinfo{person}{Christopher DuBois}, \bibinfo{person}{Alice~X Zheng}, {and} \bibinfo{person}{Martin Ester}.} \bibinfo{year}{2016}\natexlab{}.
\newblock \showarticletitle{Collaborative denoising auto-encoders for top-n recommender systems}. In \bibinfo{booktitle}{\emph{Proceedings of the ninth ACM international conference on web search and data mining}}. \bibinfo{pages}{153--162}.
\newblock


\bibitem[Xie et~al\mbox{.}(2021)]%
        {xie2021crystal}
\bibfield{author}{\bibinfo{person}{Tian Xie}, \bibinfo{person}{Xiang Fu}, \bibinfo{person}{Octavian-Eugen Ganea}, \bibinfo{person}{Regina Barzilay}, {and} \bibinfo{person}{Tommi Jaakkola}.} \bibinfo{year}{2021}\natexlab{}.
\newblock \showarticletitle{Crystal diffusion variational autoencoder for periodic material generation}.
\newblock \bibinfo{journal}{\emph{arXiv preprint arXiv:2110.06197}} (\bibinfo{year}{2021}).
\newblock


\bibitem[Xu et~al\mbox{.}(2022)]%
        {xu2022geodiff}
\bibfield{author}{\bibinfo{person}{Minkai Xu}, \bibinfo{person}{Lantao Yu}, \bibinfo{person}{Yang Song}, \bibinfo{person}{Chence Shi}, \bibinfo{person}{Stefano Ermon}, {and} \bibinfo{person}{Jian Tang}.} \bibinfo{year}{2022}\natexlab{}.
\newblock \showarticletitle{Geodiff: A geometric diffusion model for molecular conformation generation}.
\newblock \bibinfo{journal}{\emph{arXiv preprint arXiv:2203.02923}} (\bibinfo{year}{2022}).
\newblock


\bibitem[Yang et~al\mbox{.}(2024)]%
        {yang2024generate}
\bibfield{author}{\bibinfo{person}{Zhengyi Yang}, \bibinfo{person}{Jiancan Wu}, \bibinfo{person}{Zhicai Wang}, \bibinfo{person}{Xiang Wang}, \bibinfo{person}{Yancheng Yuan}, {and} \bibinfo{person}{Xiangnan He}.} \bibinfo{year}{2024}\natexlab{}.
\newblock \showarticletitle{Generate what you prefer: Reshaping sequential recommendation via guided diffusion}.
\newblock \bibinfo{journal}{\emph{Advances in Neural Information Processing Systems}}.
\newblock


\bibitem[Yi et~al\mbox{.}(2024)]%
        {yi2024directional}
\bibfield{author}{\bibinfo{person}{Zixuan Yi}, \bibinfo{person}{Xi Wang}, {and} \bibinfo{person}{Iadh Ounis}.} \bibinfo{year}{2024}\natexlab{}.
\newblock \showarticletitle{A Directional Diffusion Graph Transformer for Recommendation}.
\newblock \bibinfo{journal}{\emph{arXiv preprint arXiv:2404.03326}} (\bibinfo{year}{2024}).
\newblock


\bibitem[Yu et~al\mbox{.}(2022)]%
        {yu2022graph}
\bibfield{author}{\bibinfo{person}{Junliang Yu}, \bibinfo{person}{Hongzhi Yin}, \bibinfo{person}{Xin Xia}, \bibinfo{person}{Tong Chen}, \bibinfo{person}{Lizhen Cui}, {and} \bibinfo{person}{Quoc Viet~Hung Nguyen}.} \bibinfo{year}{2022}\natexlab{}.
\newblock \showarticletitle{Are graph augmentations necessary? simple graph contrastive learning for recommendation}. In \bibinfo{booktitle}{\emph{Proceedings of the 45th international ACM SIGIR conference on research and development in information retrieval}}. \bibinfo{pages}{1294--1303}.
\newblock


\bibitem[Zhao et~al\mbox{.}(2024)]%
        {zhao2024denoising}
\bibfield{author}{\bibinfo{person}{Jujia Zhao}, \bibinfo{person}{Wang Wenjie}, \bibinfo{person}{Yiyan Xu}, \bibinfo{person}{Teng Sun}, \bibinfo{person}{Fuli Feng}, {and} \bibinfo{person}{Tat-Seng Chua}.} \bibinfo{year}{2024}\natexlab{}.
\newblock \showarticletitle{Denoising diffusion recommender model}. In \bibinfo{booktitle}{\emph{Proceedings of the 47th International ACM SIGIR Conference on Research and Development in Information Retrieval}}. \bibinfo{pages}{1370--1379}.
\newblock


\bibitem[Zhao et~al\mbox{.}(2021)]%
        {zhao2021recbole}
\bibfield{author}{\bibinfo{person}{Wayne~Xin Zhao}, \bibinfo{person}{Shanlei Mu}, \bibinfo{person}{Yupeng Hou}, \bibinfo{person}{Zihan Lin}, \bibinfo{person}{Yushuo Chen}, \bibinfo{person}{Xingyu Pan}, \bibinfo{person}{Kaiyuan Li}, \bibinfo{person}{Yujie Lu}, \bibinfo{person}{Hui Wang}, \bibinfo{person}{Changxin Tian}, {et~al\mbox{.}}} \bibinfo{year}{2021}\natexlab{}.
\newblock \showarticletitle{Recbole: Towards a unified, comprehensive and efficient framework for recommendation algorithms}. In \bibinfo{booktitle}{\emph{proceedings of the 30th acm international conference on information \& knowledge management}}. \bibinfo{pages}{4653--4664}.
\newblock


\bibitem[Zhu et~al\mbox{.}(2024)]%
        {zhu2024graph}
\bibfield{author}{\bibinfo{person}{Yunqin Zhu}, \bibinfo{person}{Chao Wang}, \bibinfo{person}{Qi Zhang}, {and} \bibinfo{person}{Hui Xiong}.} \bibinfo{year}{2024}\natexlab{}.
\newblock \showarticletitle{Graph signal diffusion model for collaborative filtering}. In \bibinfo{booktitle}{\emph{Proceedings of the 47th International ACM SIGIR Conference on Research and Development in Information Retrieval}}.
\newblock


\end{thebibliography}

\appendix
\section{Appendix}
\begin{algorithm}[H]
	\caption{\textbf{Inference with C\ours}}  
	\begin{algorithmic}[1]
		\Require Trained flow model $f_\theta$ with parameters $\theta$, observed user interactions $X$, number of discretization steps $N$
        \State Set the starting step $s$ 
        \State Initialize $X_t \gets X$
        \For{$i = s \text{ to } N-2$}
        	\State Set the current step $t \gets t_i$
            \State Predict $\hat{X} = f_\theta(X_t, t)$ 
            \State Compute $v_t \gets (\hat{X} - X_t) / (1 - t)$
            \State Update $X_t \gets X_t + v_t / N$
        \EndFor
        \State Predict $\hat X = f_\theta(X_t, t)$ in the last step $t = t_{N-1}$
        \Ensure Predicted interactions $\hat X$
	\end{algorithmic}
    \label{alg:inference_cflowcf}
\end{algorithm}

\section{Continuous Flow-Based C\ours}
\label{sec:cflowcf}
We propose a continuous flow-based variant of \ours, referred to as C\ours, for collaborative filtering under the flexible \ours framework. While this variant retains the benefits of the behavior-guided prior, it operates in a continuous state space, diverging from the discrete nature of the interaction matrix in the original \ours. It has competitive performance compared to our VAE and diffusion-based baseline methods but introduces certain limitations.

C\ours leverages the behavior-guided prior as described in Equation~\ref{eq:sample_x0}, but it also allows for an alternative initialization strategy by directly utilizing global item frequencies as the source sample in the continuous state space:
\begin{equation*}
X_0 = \mathbf 1_{|\mathcal U|} \otimes \mathbf f \in \mathbb R^{|\mathcal U| \times |\mathcal I|}.
\end{equation*}
This initialization captures the underlying user behavior patterns but operates entirely in a continuous domain, which contrasts with the discrete formulation of the original \ours model.

To construct the flow, C\ours employs linear interpolation to generate a set of intermediate states:
\begin{equation*}
X_t = t X_1 + (1 - t) X_0.
\end{equation*}
The corresponding ground truth vector field $u$ is derived as an ordinary differential equation:
\begin{equation*}
u_t(X_t \mid X_1) = \frac{\mathrm d X_t}{\mathrm d t} = X_1 - X_0,
\end{equation*}
and by substituting $X_0$, we have:
\begin{equation*}
u_t(X_t \mid X_1) = \frac{X_1 - X_t}{1 - t}.
\end{equation*}
Similarly, we can derive the predicted vector field $v$:
\begin{equation*}
v_t(X_t) = \frac{\hat X_1 - X_t}{1 - t}.
\end{equation*}
Since all interpolations $X_t$ reside in the continuous state space, C\ours directly models these interpolations rather than their expectations $\mathbb E[X_t]$. The learning objective follows from Equation~\ref{eq:cfm_loss}:
\begin{equation*}
\mathcal L_t = \mathbb{E}_{t, X_1, X_t} \left[\|\frac{\hat X_1 - X_t}{1 - t} - \frac{X_1 - X_t}{1 - t}\|^2 \right],
\end{equation*}
which is further simplified to:
\begin{equation*}
\mathcal{L}_t = \mathbb{E}_{t, X_1, X_t} \left[\|\hat X_1 - X_1\|^2 \right].
\end{equation*}
This formulation results in a training process that is largely analogous to that of our original \ours.

During inference, C\ours directly utilizes the predicted vector field $v$ to forecast future interactions since it discards the discrete flow framework. The inference procedure is detailed in Algorithm~\ref{alg:inference_cflowcf}.

Although under the general framework of \ours, C\ours exhibits some limitations. By operating in a continuous state space, C\ours inherently disregards the binary nature of the interaction matrix, which is a fundamental characteristic of implicit feedback in recommendation tasks. The departure from our proposed discrete flow framework results in a misalignment with the underlying discrete data structure and a reduced recommendation performance, as demonstrated in our experimental results in Table~\ref{tab:ablation}.

\section{Details of the Baselines}
\label{sec:baseline}
\begin{itemize}
[leftmargin=*,noitemsep,topsep=1.5pt]
    \item \textbf{MF-BPR}~\cite{rendle2012bpr}: A matrix factorization model optimized with Bayesian Personalized Ranking (BPR) for implicit feedback.
    \item \textbf{LightGCN}~\cite{he2020lightgcn}: A collaborative filtering model that simplifies graph convolution networks with neighborhood aggregation to learn user and item representations effectively.
    \item \textbf{SGL}~\cite{wujc2021self}: A GNN-based model that leverages contrastive learning with graph augmentations on graph structures. We perform edge dropout for graph augmentation by default.
    \item \textbf{CDAE}~\cite{wu2016collaborative}: A collaborative DAE model that handles sparse and noisy interactions for top-N recommendation.
    \item \textbf{Mult-DAE}~\cite{liang2018variational}: A DAE model that uses multinomial likelihood and a point estimate approach without Bayesian regularization.
    \item \textbf{Mult-VAE}~\cite{liang2018variational}: A VAE model that uses a multinomial likelihood and variational Bayesian inference to model implicit feedback.
    \item \textbf{MacridVAE}~\cite{ma2019learning}: A VAE model disentangles user intentions and preferences through macro- and micro-disentanglement.
    \item \textbf{RecVAE}~\cite{shenbin2020recvae}: A model based on Mult-VAE with a composite prior distribution that improves reconstruction accuracy.
    \item \textbf{DiffRec}~\cite{wang2023diffusion}: A diffusion-based model inferring user preferences by modeling interaction probabilities in a denoising manner.
\end{itemize}

\section{Implementation Details}
\label{sec:implementation}
We use the Adam optimizer~\cite{kingma2014adam} with a learning rate of 0.001, no weight decay, and a batch size of 4096. The hidden sizes of the MLP in the flow model are set to [300, 300] for ML-1M and ML-20M, and [600] for Amazon-Beauty. The initial dropout rate of the MLP in \ours is fixed at 0, while for other generative CF baseline models, it is tuned within [0, 0.5].
We run and tune our baselines using the widely used recommendation library RecBole~\cite{zhao2021recbole} and report the best results to ensure a fair comparison. The best hyperparameters are selected based on NDCG@10 on the validation sets.

\end{document}